\documentclass[11pt,a4paper]{article}
\pdfoutput=1

\usepackage{jheppub}

\usepackage{colortbl}
\usepackage{epsfig,multicol,bbm}
\usepackage{amsmath,amsfonts,amssymb,longtable,mathtools}
\usepackage{array}
\usepackage{graphicx}
\usepackage{subfig}
\usepackage{enumerate}
\usepackage{color}
\usepackage{placeins}
\usepackage{longtable,booktabs,caption}
\usepackage{changepage}
\usepackage{float}
\usepackage[makeroom]{cancel}
  \usepackage{latexsym}
  \usepackage{epsf}
  \usepackage{amssymb}
  \usepackage{graphicx}
  \usepackage{amsmath}
  \usepackage{amsmath,amssymb,amsthm}
  \usepackage{verbatim}
  \usepackage{hyperref}

\newcommand \ms {M_{\odot}}
\begin{document}

\title{Primordial Black Hole Formation in Non-Standard Post-Inflationary Epochs}

% Authors, for the paper (add full first names)
\author{Sukannya Bhattacharya\footnote{email: sukannya.bhattacharya@unipd.it}$^{a,b}$}
%%\longauthorlist{yes}
%
%% MDPI internal command: Authors, for metadata in PDF
%\AuthorNames{Sukannya Bhattacharya}

% MDPI internal command: Authors, for citation in the left column
%\AuthorCitation{Bhattacharya, S.}
% If this is a Chicago style journal: Lastname, Firstname, Firstname Lastname, and Firstname Lastname.

% Affiliations / Addresses (Add [1] after \address if there is only one affiliation.)
\affiliation{$^{a}$Dipartimento di Fisica e Astronomia, Universit\`a degli Studi di Padova, Via Marzolo 8, 35131 Padova, Italy\\
$^{b}$Istituto Nazionale di Fisica Nucleare (INFN), Sezione di Padova, Via Marzolo 8, 35131 Padova, Italy}

% Contact information of the corresponding author
%\corres{Correspondence: }

\abstract{When large overdensities gravitationally collapse in the early universe, they lead to primordial black holes (PBH). Depending on the exact model of inflation leading to necessary large perturbations at scales much smaller than scales probed at the Cosmic Microwave Background (CMB) surveys, PBHs of masses $\lesssim$$10^3 \ms$ are formed sometime between the end of inflation and nucleosynthesis. However, the lack of a direct probe for the exact expansion history of the universe in this duration introduces uncertainties in the PBH formation process. The presence of alternate cosmological evolution for some duration after inflation affects the relation between (i) PBH mass and the scale of the collapsing overdensity; and (ii) PBH abundance and amplitude of the overdensities. In this review, the non-standard cosmological epochs relevant for a difference in PBH production are motivated and discussed. The importance of developing the framework of PBH formation in non-standard epochs is discussed from a phenomenological point of view, with particular emphasis on the advances in gravitational wave (GW) phenomenology, since abundant PBHs are always accompanied by large induced GWs. PBH formation in general non-standard epochs is also reviewed including the mathematical formalism. Specific examples, such as PBH formation in a kinetic energy dominated epoch and an early matter dominated epoch, are discussed with figures showing higher PBH abundances as compared to the production in standard radiation domination. 
}
%\keyword{Primordial Black Holes, Early Universe, Inflation } 
\maketitle
%\makeindex

%\begin{document}
\section{Introduction}
\label{intro}
Primordial black holes (PBHs) have taken a seat at the forefront of contemporary research in cosmology. PBHs are nonrelativistic and effectively collisionless, properties which make them viable candidates for dark matter (DM)~\cite{Hawking:1971ei,Carr:1974nx,Carr:1975qj,Carr:2021bzv,Carr:2020xqk}. With the recent observations of binary black hole systems by LIGO/Virgo surveys~\cite{LIGOScientific:2016aoc,LIGOScientific:2016sjg,LIGOScientific:2016dsl,LIGOScientific:2016wyt,LIGOScientific:2017bnn,LIGOScientific:2017vox,LIGOScientific:2017ycc}, there is a possibility that some of the black holes observed are not astrophysical, but of primordial origin~\cite{Fernandez:2019kyb}. Since PBHs are formed in the early universe, inspecting them phenomenologically can convey a better understanding of the universe at very high energies.
PBHs can have masses spanning over a huge range from $\sim$$10^{15}$ gm  to $\sim$$10M_{\odot}$, where percent level contributions of PBH to the total DM abundance are still not ruled out by observations~\cite{Carr:2020gox}. Therefore, several types of experiments can be used for constraining PBHs, ranging from galactic and extragalactic $\gamma$-ray detectors relevant for light PBHs to lensing surveys and binary merger observations for the heavy ones~\cite{Carr:2020gox,Green:2020jor}.  %Depending on the mechanism of production, PBHs can also have spin, which also adds to the interest since current observations of black hole mergers also measure the spin of the black holes very precisely. 

In the context of early universe cosmology, understanding the reason and mechanism for the production of PBHs is crucial~\cite{Green:2014faa,Young:2014ana,Bloomfield:2015ila,Kuhnel:2015vtw,Georg:2016yxa,Young:2016mxm,Young:2019yug,Villanueva-Domingo:2021spv,Gow:2021tpe}. PBHs can be formed due to different mechanisms, such as the collapse of density perturbations which originate from single field~\cite{Ivanov:1994pa,Yokoyama:1998pt,Garcia-Bellido:2017mdw,Ballesteros:2017fsr,Hertzberg:2017dkh,Kinney:2005vj,Germani:2017bcs,Pattison:2017mbe,Ezquiaga:2018gbw,Biagetti:2018pjj,Stewart:1997wg,Kohri:2007qn,Alabidi:2009bk} or multi-field models~\cite{Randall:1995dj,Garcia-Bellido:1996mdl,Kawasaki:1997ju,Clesse:2015wea,Lyth:2001nq,Kawasaki:2012wr,Kohri:2012yw,Yokoyama:1995ex,Bhattacharya:2022fze} of inflation, from bubble collisions~\cite{Crawford:1982yz,Hawking:1982ga,La:1989st,Moss:1994iq,Konoplich:1999qq,1998AstL...24..413K,Sato:1980yn,Guth:1980zm}, collapse of cosmic strings~\cite{Planck:2013mgr,Blanco-Pillado:2017rnf,Hawking:1987bn,Polnarev:1988dh,Hansen:1999su,Hogan:1984zb,Nagasawa:2005hv,James-Turner:2019ssu,Caldwell:1991jj,MacGibbon:1997pu,Jenkins:2020ctp,Helfer:2018qgv,Matsuda:2005ez,Lake:2009nq} or domain walls~\cite{Rubin:2001yw,Dokuchaev:2004kr,Khlopov:2000js,Ge:2019ihf,Garriga:2015fdk,Deng:2016vzb,Deng:2017uwc,Liu:2019lul,Kopp:2010sh,Harada:2004pe} or scalar fields~\cite{Cotner:2016cvr,Cotner:2017tir,Cotner:2018vug,Cotner:2019ykd}, etc. Each of these mechanisms leads to a specific mass spectrum of the PBH produced, which leads to the relative abundance of PBHs as DM, a quantity that can be checked with observational~bounds. 

In particular, PBH formation from the collapse of large overdensities is highly interesting since these overdensities in the early universe can be linked to the primordial quantum fluctuations produced during inflation~\cite{Carr:1974nx,Carr:1975qj,Young:2019yug,Kuhnel:2015vtw}. Scalar fluctuations are produced at all scales, which exit the horizon when the universe expands quasi-exponentially during inflation. At the end of inflation, these fluctuations re-enter the horizon one by one, become classical density fluctuations and grow. If large overdensities are present, they can gravitationally collapse with a certain probability and form PBHs. The superhorizon behaviour of the fluctuations depends on the model of inflation, whereas their subhorizon growth in the post-inflationary epochs depends on the energy density driving that epoch. Therefore, given a model of inflation that can produce large scalar fluctuations, the formation of PBH depends on the dominant component for the energy density at the time of the collapse. 

In the standard picture, at the end of inflation, reheating takes place either instantaneously or slowly during which the universe becomes populated with relativistic degrees of freedom (dof). At the end of reheating, these relativistic species start dominating the energy density of the universe, thus marking the onset of radiation domination (RD). The physics of reheating and preheating, although theoretically developed~\cite{Bassett:2005xm,Frolov:2010sz,Allahverdi:2010xz,Amin:2014eta,Lozanov:2019jxc},
 cannot be probed independently as these epochs are largely dependent on the model of inflation. The span in energy densities from the end of inflation ($\sim$$10^{16}$ GeV) and big bang nucleosynthesis (BBN) ($T_{\rm BBN}\sim 5$ MeV) is huge $\sim$$\mathcal{O}(10^{19})$, and is not accessible to direct observational probes. The observed abundance of light elements requires the universe to be RD at least by the time of BBN. Therefore, there is a certain possibility that the evolution in the history of the universe deviated once or multiple times from this simple picture of RD in this range. 

%the early universe becomes radiation dominated (RD) at the end of inflation and  reheating. Constraints on the abundance of light elements demand that (?) the universe becane RD at least by the time of big bang nucleosynthesis (BBN) at temperature $T\sim 5$ MeV. However, the physics of reheating and possible preheating is not clear and typically depends on the model of inflation. Moreover, there is no direct observational probe yet to know about the exact progression from the end of inflation ($\sim 10^{16}$ GeV) and BBN, which can span an order $\sim 10^{19}$ in the ratios of the energy densities. There can be simple or complicated, one-time or multiple deviations from the standard RD picture during this time. 

Since the PBH formation process and the resulting abundance depend crucially on the overdensity at the time of collapse as well as the evolution of relative energy density of PBH and background, it is of immense importance to investigate the scenario when PBH is formed in non-standard post-inflationary epochs~\cite{Allahverdi:2020bys,Carr:2018nkm}. There can be several reasons which may give rise to such a non-standard evolution, e.g., prolonged reheating, a heavy scalar field that can dominate the energy density for some time and then reheat the universe again, a sterile field dominating the energy density, the kinetic energy of a scalar field dominating the energy budget, etc.. These various scenarios have been discussed in detail in Section~\ref{nonst}, and PBH formation has been analysed for a general non-standard epoch and a few well-motivated examples in Sections~\ref{PBHformw} and~\ref{PBHspecific}.

In the realm of contemporary research on primordial cosmology, theoretical model building for the early universe goes hand in hand with observational data. In this aspect, PBHs provide a uniquely interesting indirect probe towards the early universe at high energy scales combining the details of inflationary dynamics and the post-inflationary evolution of the universe. The collapse of large density perturbations originating from inflationary scalar fluctuations is one of the most studied mechanisms to generate PBHs. On one hand, the volume of literature is growing to realise models of inflation with single or multiple fields in simple or exquisite settings such as in the presence of a thermal bath (warm inflation), turns in the field space (multi-field inflation), non-trivial gravitational and derivative couplings, non-canonical kinetic terms, etc., which can predict CMB consistent amplitude and spectral index for the scalar perturbations at the CMB scales, and simultaneously include growth of fluctuations and therefore blue-tilted/peaked power spectra at smaller scales. If the small-scale inflationary power spectrum is large enough ($\sim$$0.02$), it can lead to a copious amount of PBH formation in the RD epoch after the end of inflation.

However, on the other hand, various scenarios of alternate cosmological evolutions after inflation are being proposed, to explain, for example, the post-inflationary fate of the inflaton/spectator fields/moduli fields or to incorporate additional dof which dominate the energy density for some time, etc. For a given inflationary power spectrum $\mathcal{P}_{\zeta}(k)$, with large amplitude at small scales, the mass spectrum of PBH is affected if they are formed in such non-standard epochs. This review attempts to discuss the possible reasons behind the occurrence of such non-standard post-inflationary epochs as well as their effects on the resulting abundance and relevant mass range for PBH, with attention to how the basic contributory quantities are affected.

This review is structured as follows: in Section~\ref{importance}, the necessity for PBH analysis in non-standard postinflationary epochs has been motivated. In Section~\ref{nonst}, possible and relevant non-standard epochs have been discussed. In Section~\ref{reason}, a clear picture is provided for the horizon exit and re-entry of the inflationary fluctuations. In Section~\ref{analysis}, mathematics to estimate PBH mass, mass spectra and abundance has been developed. This has been carried out in two parts, one for a general non-standard epoch with nonzero pressure, and one for a matter dominated epoch. In the same section, having been introduced to the components that affect PBH formation and abundance, the effects of different contributors and different methods to estimate them have been discussed. A few specifically interesting examples of non-standard epochs have been discussed in Section~\ref{PBHspecific}, with results shown for two particular forms of the primordial power spectrum. In Section~\ref{conc}, discussions on the current status and future prospects have been made. In this review, the reduced Planck mass is denoted as $M_{\rm P}=2.44\times 10^{18}$ GeV, and the solar mass is denoted as $\ms = 2\times 10^{33} {\rm gm}\simeq 10^{57}$ GeV.
%\section{Structure In addition, Scope Of The Review}
%\subsection*{Other useful references}
%An interested reader may consult the following references for detailed picture on the topic of this review as well as related topics on PBH.\\
%(i) PBH formation and abundance:~\cite{Green:2014faa,Young:2014ana,Bloomfield:2015ila,Kuhnel:2015vtw,Georg:2016yxa,Young:2016mxm,Young:2019yug,Villanueva-Domingo:2021spv,Gow:2021tpe};\\
%(ii) Constraints on PBH abundance:~\cite{Carr:2020gox,Green:2020jor};\\
%(iii) non-standard epochs in the early universe:~\cite{Allahverdi:2020bys,Carr:2018nkm}.

%----------------------------------------------------------------------------------------------------------------

\section{Importance in Current Phenomenology}
\label{importance}
Several types of observational and experimental data now constrain a significant part of the PBH parameter space. These constraints are expected to evolve in the near future with the prospect of additional data and improved analysis. PBHs evaporate on a timescale $t_{\rm ev}=5120\pi G^2M^3/(\hslash c^4)$ via Hawking radiation, and therefore PBHs of mass lower than $M\simeq 5\times 10^{14}~{\rm g}\simeq 2.5\times 10^{-19}\ms$ have completely evaporated by now \cite{Hawking:1974rv}. Slightly heavier PBHs have not completely evaporated yet and may radiate gamma-ray photons, neutrinos, gravitons and other massive particles at different stages of evaporation. Therefore, by constraining the injection of photons and neutrinos in the (extra-)galactic medium using Voyager data, extra-galactic radiation background, SPI/INTEGRAL observations, etc.~\cite{Churazov:2010wy,Siegert:2016ijv,Laha:2019ssq,Bays:2011si,Collaboration:2011jza,Agostini:2019yuq,Dasgupta:2019cae,Laha:2020ivk}, limits can be put on the abundance of light PBHs with $M \lesssim 10^{-17}\ms$. CMB anisotropies and abundance of light elements at the time of BBN due to the energy decomposition in the background by the evaporation products from the black holes~\cite{Acharya:2020jbv} can constrain PBHs for masses $M \geq 5.5\times 10^{-21} \ms$ and $M \simeq 10^{-22} - 10^{-21}\ms$, respectively. PBHs in the mass range $10^{-11}\ms<M<10^{-1}\ms$ are constrained by their gravitational lensing of light rays from distant stars. Observation of the stars in the M31 galaxy by the HSC telescope, the EROS and OGLE survey together now rule out the contribution of PBH towards total DM density above 1--10\% in this mass range \cite{Smyth:2019whb,Tisserand:2006zx,Niikura:2017zjd,Niikura:2019kqi}. The caustic crossing event for the star Icarus or MACS J1149LS1 and the resultant strong lensing has been used to place constraints on compact objects in the range $10^{-5}\ms<M\lesssim 10^3\ms$~\cite{Oguri:2017ock}. The GW detections by the LIGO/Virgo collaboration put an upper bound on the total PBH abundance in the mass region $0.2\ms<M<300\ms$, assuming that the observed binary BH mergers are PBH mergers in the early or late universe~\cite{Ali-Haimoud:2017rtz,Bird:2016dcv,Sasaki:2016jop,Cholis:2016kqi,Clesse:2016vqa,Raccanelli:2016cud,Kovetz:2017rvv,Authors:2019qbw,Kavanagh:2018ggo,DeLuca:2020qqa,Wang:2016ana}. Finally, the radiation from the accreted gas around PBHs of mass $M \gtrsim 100\ms$ affects the spectrum and the anisotropies of the CMB~\cite{carr1981pregalactic,Ricotti:2007au,Serpico:2020ehh}. 

\begin{figure}[H]
\includegraphics[width=0.9\textwidth]{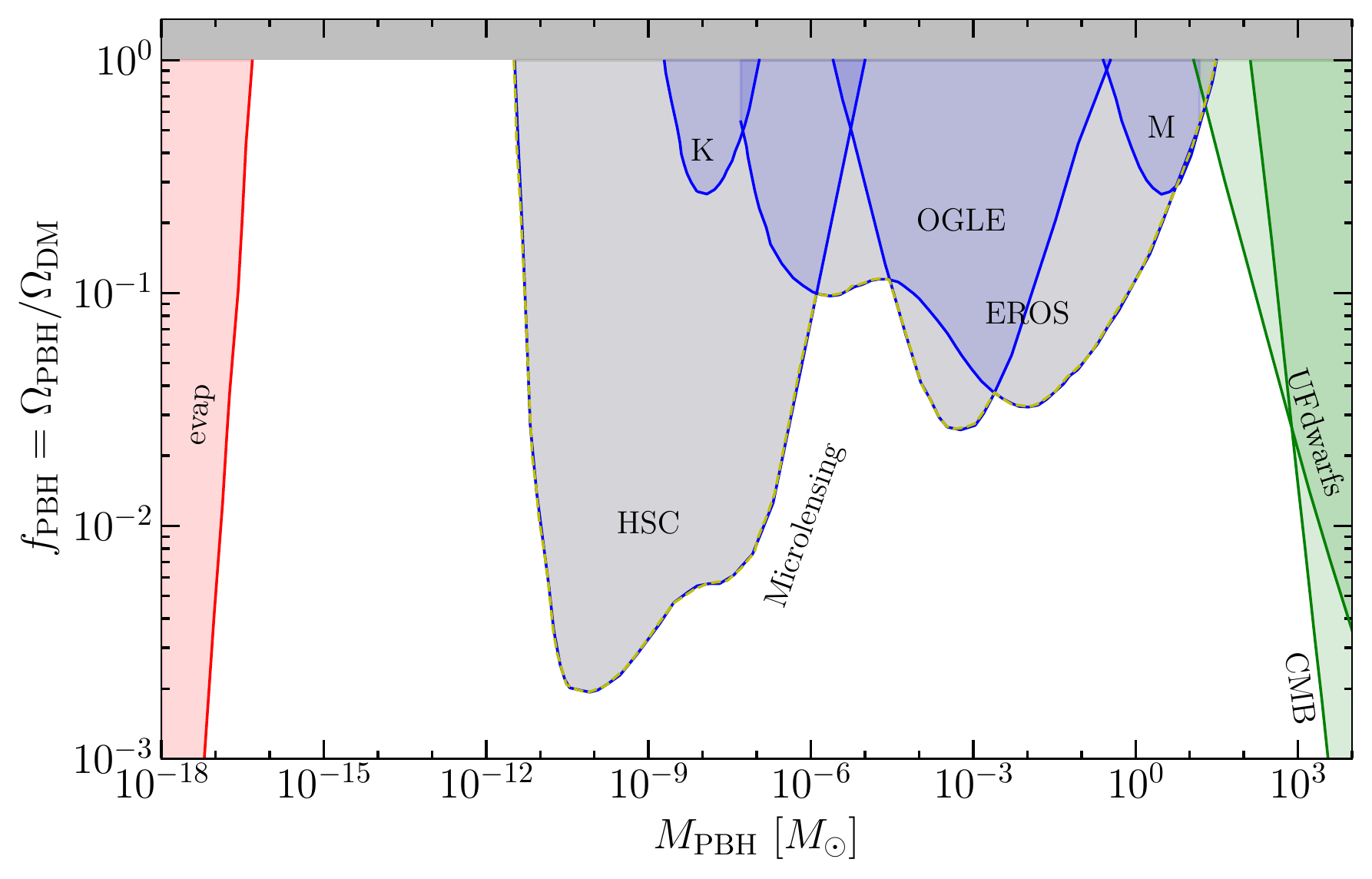}
\caption{Examples of bounds on the abundance of PBH are plotted using~\cite{bradley_j_kavanagh_2019_3538999}. `evap' signifies bounds from PBH evaporation~\cite{Lehmann:2018ejc} via Hawking radiation; HSC~\cite{Niikura:2017zjd}, Kepler (K)~\cite{Griest:2013aaa}, EROS~\cite{Tisserand:2006zx}. MACHO~\cite{Macho:2000nvd} and OGLE~\cite{Niikura:2019kqi} signify the bounds from microlensing (yellow dashed line enveloping the blue and grey shaded regions); `UFdwarfs' signifies bounds from ultra-faint dwarf galaxies~\cite{Brandt:2016aco}; `CMB' signifies bounds from Compton drag and Compton cooling of CMB photons~\cite{Ali-Haimoud:2016mbv}. }
\label{pbhbounds}
\end{figure}

In Figure~\ref{pbhbounds}, a few of these bounds from several types of observations are shown for monochromatic PBH mass spectrum~\cite{bradley_j_kavanagh_2019_3538999}. Clearly, the only remaining window for PBH to form $100\%$ of dark matter is $10^{-16}\ms<M\lesssim 10^{-12}\ms$. However, there are bounds from the capture of PBH by neutron stars at the dense core of a globular cluster~\cite{Capela:2013yf} and, from the shape of the observed distribution of white dwarfs~\cite{Graham:2015apa}, can put constraints on this mass range as well. More stringent constraints in this mass range are expected to come from the future observation of the stochastic background of induced GW in upcoming surveys such as LISA~\cite{LISA:2017pwj,Kaiser:2020tlg,Barausse:2020rsu,LISACosmologyWorkingGroup:2022kbp} and DECIGO~\cite{Seto:2001qf,Yagi:2011wg,Kawamura:2020pcg}. 
%If the recent observation of  
%a common power-law spectrum by NANOGrav 12.5 year data are confirmed to be the signature of stochastic background of GWs then PBH abundance around $M\sim 1-10 \ms$ can be further constrained.

The abundance and masses of the PBH produced with a certain mechanism depend on the details of the underlying model. In the case where the large overdensities collapsing into PBH result from primordial inflationary fluctuations, there is a direct relation between (i) the scales $\sim$$1/k$ ($k$ is the wavenumber) for which the primordial perturbations $\zeta$ are large, and the PBH mass $M$; (ii) amplitude of the enhanced fluctuations (amplitude of the primordial power spectrum $\mathcal{P}_{\zeta}(k)$) and the PBH mass spectrum $\psi (M)$, as well as the total PBH abundance $f_{\rm PBH}$. Evidently, with the current constraints on PBH abundance, it is extremely important to confront relevant models of inflation, which can generate large $\mathcal{P}_{\zeta}(k)$ at small scales, with PBH phenomenology. This way, constraints on PBH abundance can put bounds on the inflationary fluctuations. However, the observational bounds discussed above are typically given for a monochromatic PBH mass spectrum, which is not the case when one analyses PBH formation from realistic models of inflation. In~\cite{Carr:2017jsz}, a method was developed to generate bounds on the extended PBH mass spectra, once the bounds for the monochromatic spectrum are known. Once the extended PBH mass spectrum obtained from an inflation model is treated with these modified observational bounds, proper upper limits on $\mathcal{P}_{\zeta}(k)$ can be imposed. 

However, this statement is not exact if PBHs are formed in a non-standard epoch of the unknown equation of state (EoS) and duration in the post-inflationary universe. In fact, PBH formation in a non-standard epoch with EoS $w$ lasting from temperature $T_*$ to $T_{\rm RD}$ affects the one-to-one relationships between $M$ and $k$ and between $\psi (M)$ and $\mathcal{P}_{\zeta}(k)$. The dynamics of PBH formation obtains contributions from $w$ and $T_{\rm RD}$\footnote{$T_*$ does not affect PBH formation when we assume that the formation process begins during the $w$-dominated epoch, since the energy fraction contained in PBH at the time of formation depends only on the temperature at formation.}. These exact $w$ and $T_{\rm RD}$-dependent relations are developed in Section~\ref{PBHformw} in this review.

Large fluctuations required for abundant PBH are usually accompanied by large non-Gaussianities.
The effect of primordial non-Gaussianity in the PBH abundance has also been discussed in literature~\cite{Young_2013,Young_2016,Franciolini_2018,Luca_2019,Yoo_2019,Kehagias_2019}. 
The non-Gaussianity parameters, such as $f_{\rm NL}$, are weekly constrained even at the CMB scales by Planck 2018~\cite{Planck:2018jri,Aghanim:2018eyx}. However, constraints on PBH abundance while accounting for primordial non-Gaussianities can provide new bounds on these parameters at small scales, which sheds light on the underlying mechanism at small scales of inflation.

If the primordial spectrum contains certain features, then they can be translated into the PBH mass spectrum. For example, resonant oscillations around the peak of $\mathcal{P}_{\zeta}(k)$, which is ubiquitously seen in multi-field models of inflation with turns in the field \mbox{space~\cite{Flauger:2009ab,Flauger:2014ana,Gao:2015aba,Fumagalli:2020adf,Fumagalli:2020nvq,Braglia:2020taf,Fumagalli:2021cel,Fumagalli:2021dtd,Bhattacharya:2022fze},} can be carried over to explicit oscillations in $\psi (M)$, depending on the detailed reason behind the resonant oscillations in $\mathcal{P}_{\zeta}(k)$ in the first place. 
%In some cases of multi-field inflation, observational bounds on the coupling(s) between the inflaton and accompanying field(s) can contribute to PBH abundance, which, when treated with observations, may be crucial to understand the field theoretical description of inflation. 

However, in the presence of a non-standard epoch, all of these constraints on PBH getting translated into bounds on the small scale $\mathcal{P}_{\zeta}(k)$ require inputs for the exact values of $w$ and $T_{\rm RD}$.
In the presence of a non-standard epoch, the PBH abundance obtains inputs from both the inflationary paradigm and the $w$-dominated epoch. Mathematically speaking, both of these inputs can affect the final PBH abundance by orders of magnitude since $\mathcal{P}_{\zeta}(k)$ appears as an exponent (see Equation~\eqref{probdel}), and $w$ appears in the powers of the $M$ and $T_{\rm RD}$ (see Equation~\eqref{Mkexact}). Therefore, considering the progress in lowering observational bounds on PBH, it is necessary and timely to not only model the inflationary paradigm but also investigate the possibility of a non-standard post-inflationary epoch.

Another extremely interesting avenue is induced gravitational waves (IGW) that can be combined with the PBH phenomenology to provide a better understanding of the early universe at small scales. In the second and higher orders of perturbation theory, scalar and tensor perturbations are coupled. Therefore, adiabatic perturbations source higher order tensor fluctuations~\cite{Ananda:2006af,Baumann:2007zm,Kohri:2018awv,Espinosa:2018eve} (for a recent review, see \cite{Domenech:2021ztg}), which are subdominant with respect to the first order tensor modes for simple slow-roll models of inflation with red-tilted adiabatic power spectra. However, an enhanced $\mathcal{P}_{\zeta}(k)$ can lead to a large induced tensor power spectrum and therefore a large spectrum of IGW. Such IGWs are primordial in nature and appear as stochastic backgrounds today. With the prospect of ground/space-based interferometric detectors and pulsar timing arrays, the growing interest in using IGW as a probe for the early universe is promising to have a detailed understanding of the primordial fluctuations.
At large scales, the scalar fluctuations are tightly constrained by CMB observations and thus result in IGWs of tiny amplitude.
However, models where the scalar fluctuations at small scales are significantly enhanced, such as those leading to large PBH abundance, can lead to large IGW spectra  simultaneously.
  %%%
%multi-field inflation scenarios enhanced  scalar power spectrum with resonant features at small scales as mentioned above,  may give rise to  potentially detectable IGWs with interesting features 
%\cite{Fumagalli:2020nvq,Braglia:2020taf,Fumagalli:2021cel,Witkowski:2021raz,Fumagalli:2021mpc,Fumagalli:2021dtd}.

Therefore, the scenario of abundant PBH formation is always accompanied with large IGWs, but both of the dynamics depend on the epoch of collapse and the epoch of the IGW sourcing of GW from scalar modes, respectively. 
The frequency $f$ of the IGW depends on the mode $k$ entering the horizon at the post-inflationary time when the GW is sourced. If PBHs are formed in the radiation dominated (RD) epoch, then the wavenumber $k$ entering the horizon, PBH mass $M$ and frequency $f$ of the IGW are related via the following relation:
\begin{equation}
\bigg(\frac{M}{\ms}\bigg)^{-1/2}\simeq \frac{k}{2\times 10^{14}~{\rm Mpc}^{-1}}=\frac{f}{0.3~ {\rm Hz}}.
\label{MkfRD}
\end{equation} 

However, when they are produced in a general $w$-dominated epoch, then this relation is modified as\footnote{The numerical factor in the denominator of the second equality of~\eqref{MkfwD} arises from $(4\pi\gamma C(w))^{\frac{1+3w}{3(1+w)}}$ in Equation~\eqref{Mkexactsol}, and its value is put as $2\times 10^6$ here. It can vary between (2--6)$~\times~10^6$ for $1\geq w>0$.
}:
\begin{equation}
\bigg(\frac{M}{\ms}\bigg)^{-\frac{1+3w}{3(1+w)}}\bigg(\frac{T_{\rm RD}}{\rm GeV}\bigg)^{\frac{1-3w}{3(1+w)}}\simeq \frac{k}{2\times 10^{6}~{\rm Mpc}^{-1}}=\frac{f}{3~{\rm nHz}},
\label{MkfwD}
\end{equation}

The present and proposed GW surveys span over decades in the frequency space. Pulsar timing arrays (PTAs), such as NANOGrav~\cite{Aggarwal:2018mgp,NANOGrav:2020bcs}, EPTA~\cite{Lentati:2015qwp,Shannon:2015ect,Qin:2018yhy}, etc., are sensitive in the range $10^{-9}$--$10^{-7}$ Hz, corresponding to $6\times 10^5 {~\rm Mpc}^{-1}\lesssim k \lesssim 6\times 10^7 {~\rm Mpc}^{-1}$. Ground based interferometric detectors such as LIGO/Virgo~\cite{LIGOScientific:2019vic,LIGOScientific:2016aoc,LIGOScientific:2016dsl,LIGOScientific:2017ycc}, KAGRA~\cite{Akutsu:2015hua,Haino:2020age} and ET~\cite{Maggiore:2019uih} cover the range $10$--$10^{3}$ Hz, corresponding to $6\times 10^{15} {~\rm Mpc}^{-1}\lesssim k \lesssim 6\times 10^{18} {~\rm Mpc}^{-1}$. The intermediate frequency range can be probed by LISA~\cite{LISA:2017pwj,Kaiser:2020tlg,Barausse:2020rsu,LISACosmologyWorkingGroup:2022kbp}, DECIGO~\cite{Seto:2001qf,Yagi:2011wg,Kawamura:2020pcg}, AION/MAGIS~\cite{Badurina:2019hst}, Taiji~\cite{Ruan:2018tsw}, and TianQin~\cite{TianQin:2015yph}.

With the prospect of current and upcoming GW surveys, in the optimistic scenario with positive detection of GW, the primordial $\mathcal{P}_{\zeta}(k)$ can have constraints on its amplitude and spectral index, which can help decrease the model space for inflation. However, GWs at the stochastic level may have several cosmological and astrophysical sources, and it may be challenging to recognise an IGW signal with confidence. One possible solution is to check the spectral index of the observed GW spectrum since different processes predicting stochastic GW signals usually have specific spectral signatures of the predicted  GW spectra. Even if there is no positive observation of GW, with gradually improving sensitivities of the GW surveys, stricter upper bounds on $\mathcal{P}_{\zeta}(k)$ can be provided. 

%%%new addition for arXiv version
Other than primordial scalar fluctuations, gravitational waves can also be induced by the Poisson isocurvature perturbations of very light PBHs (with $M <10^9$ g), accompanied by an early PBH-dominated ($w=0$) epoch~\cite{Papanikolaou:2020qtd,Papanikolaou:2022chm}. Light PBHs formed in standard RD or nonstandard epochs can also lead to gravitational waves via Hawking evaporation~\cite{Ireland:2023avg}.

With the possibility of observing and constraining IGWs, the PBH phenomenology is also improved, since we can obtain an even better understanding of the small scale inflationary dynamics. Starting from a model of inflation with enhanced $\mathcal{P}_{\zeta}(k)$ at small scales, one generally studies the predictions for both PBH and IGW. However, the presence of a non-standard epoch affects both of these processes. The IGW spectrum for a general $w$-dominated epoch has been developed in detail in literature~\cite{Domenech:2021ztg,Bhattacharya:2019bvk}.

Hence, even with the combined PBH-IGW analysis and phenomenology, which is of great interest to current trends in inflationary model building, if a non-standard epoch is present after inflation, then predictions and constraints must be rechecked~\cite{Bhattacharya:2020lhc}. One interesting and hopeful aspect of such a combined phenomenological study is that the relation between $k$ and $f$ is always $k=2\pi f$, independent of which epoch the IGW is sourced in. Thus, with an IGW observation, the actual peak position of $\mathcal{P}_{\zeta}(k)$ can be found, irrespective of a non-standard epoch. However, the relation between $M$ and $k$ depends on $w$ and $T_{\rm RD}$, and therefore can separately give information about the $w$-dominated epoch in a combined study. However, this is not so straightforward as the IGW spectrum and $\psi (M)$ both depend on the post-inflationary evolution, and therefore on $w$ and $T_{\rm RD}$.
\section{Non-Standard Epochs after Inflation}
\label{nonst}
There can be several scenarios leading to one or multiple epoch(s) of non-standard expansion before or after BBN. This review discusses the deviations from standard evolution only before BBN because the mass of the PBH corresponding to BBN is $M_{\rm BBN}\sim 10^3 M_{\odot}$. The PBHs of phenomenological interest, which can lead to reasonable DM abundance with several bounds from astrophysical and cosmological surveys, are in the range of $\sim$$10^{15}$ gm and $\leq 100M_{\odot}$, which form before BBN. Such post-inflationary and pre-BBN non-standard epochs can arise either from modifications of standard $\Lambda$CDM properties or the introduction of entirely new components. However, in this review, we divide them into two categories: \textit{reheating} which begins at the end of inflation and leads to standard RD either instantaneously or slowly; and \textit{ general $w$-domination}, which begins at some point during standard RD and ends by the time of BBN. 

\subsection{Reheating}
At the end of inflation, the inflaton ($\phi$) energy density needs to be transferred to the Standard Model (SM) degrees of freedom, as well as DM to commence standard RD. This intermediate epoch, named (p)reheating~\cite{Dolgov:1989us,PhysRevD.42.2491}, is governed by the shape of the inflaton potential near the minimum and the couplings of the inflaton to other fields. In case of negligible couplings, if the single field inflaton potential has the form $V(\phi) \propto \vert \phi \vert ^{2n}$ near the minimum, then the homogeneous inflaton condensate executes quasi-periodic oscillations around the minimum of $V(\phi)$ while the time-averaged equation of state (EoS) has the form~\cite{PhysRevD.28.1243}
\begin{equation}
w=\frac{n-1}{n+1}.
\label{wn-reh}
\end{equation}

The process of reheating can include perturbative and/or non-perturbative parts. Since the effective inflaton mass $m_{\rm eff}^2\equiv \frac{d^2V}{d\phi ^2}$ varies with time during the oscillations of the condensate, resonant transfer of energy from the condensate to shorter wavelength modes is possible~\cite{Kofman:1994rk,Shtanov:1994ce,Kofman:1997yn}, leading to rapid and non-adiabatic growth of short-wavelength fluctuations. The duration of the $w$-dominated epoch depends on the full shape of the inflaton potential. Quadratic behaviour of $V(\phi)$ near the minimum ($n=1$) leads to $w=0$, i.e., a matter dominated (MD) epoch, whose duration depends on the gravitational interactions of the inflaton condensate~\cite{Amin:2010xe,Amin:2010dc,Amin:2011hj,Gleiser:2011xj,Lozanov:2017hjm,Hong:2017ooe,Fukunaga:2019unq,Gleiser:1993pt,Copeland:1995fq,Kasuya:2002zs,Hindmarsh:2006ur,Amin:2010jq,Zhang:2020bec}. For a quartic form of $V(\phi)$ near the minimum, ($n=2$), $w=1/3$, i.e., a RD epoch is approached~\cite{Lozanov:2017hjm,Lozanov:2016hid}. However, typically at the end of the resonant decay of the condensate, coupling of the inflaton to other fields needs to be invoked for the inflaton to decay completely. 

In the multi-field inflation models, if the inflaton is directly coupled to other fields, then the latter have effective masses dependent on $\phi$~\cite{Bassett:2005xm,Frolov:2010sz,Allahverdi:2010xz,Amin:2014eta,Lozanov:2019jxc,Kofman:1997yn}. These couplings typically shorten the duration of the $w$-dominated epoch due to increased efficiency of decay processes. Moreover, inflaton and the other fields may have nonminimal coupling to gravity, or nontrivial field-space manifolds, which can lead to noncanonical kinetic terms that aid in the resonant decay of the inflaton. In warm inflation models~\cite{Berera:1995ie,Berera:2008ar} where the inflaton energy density dissipates to a thermal bath during inflation, the reheating process may be even more hastened, if at all necessary. 

The reheating process for both single and multi-field inflation scenarios is extremely model dependent, more so in the latter case. Therefore, a $w$-dominated epoch during reheating is also dependent on the underlying inflation model and is relevant for the PBH masses of interest only if this epoch is prolonged.

\subsection{General $w$-Dominated Epoch}
\label{genw1}
There are many well-motivated scenarios where the post-reheating universe is dominated by a particle species $\Phi$ with a general EoS $w$, so that $\rho _{\Phi}\propto a^{-3(1+w)}$. For example, an early matter dominated (EMD) epoch ($w=0$) may arise when a heavy field drives the energy density of the universe~\cite{Vilenkin:1982wt,Coughlan:1983ci,Starobinsky:1994bd,Dine:1995uk,Chung:1998rq}. A well-motivated example of this kind is an epoch dominated by moduli fields in several string inflation models~\cite{Kane:2015jia, Allahverdi:2020bys}. On the other hand, an epoch dominated by the kinetic energy density of a fast-rolling field has $w\simeq 1$. This may take place after an epoch of quintessential inflation~\cite{Peebles:1998qn,Ahmad:2019jbm}, when the inflaton field rolls down very fast from its inflaton potential towards the potential relevant for dark energy at a later stage. QCD phase transition may lead to a softening of the background energy density of RD, i.e., EoS becomes $w<1/3$ for a small duration. These three special cases will be discussed explicitly in reference to PBH formation in later sections. 

More general values of $w$ are possible when a scalar field oscillates with a particular potential form~\cite{Choi:1999xn,Gardner:2004in,DEramo:2017gpl,DiMarco:2018bnw}, in braneworld cosmologies~\cite{Okada:2004nc,Meehan:2014bya}, scalar-tensor theories of gravity~\cite{Catena:2004ba,Dutta:2016htz}, etc. Particularly, stiff EoS $1/3<w\leq 1$ may arise when a sterile field enters the post-inflationary phase with a dominant energy contribution~\cite{DiMarco:2018bnw}. 

The onset of such non-standard epochs at temperature $T_*$ can be determined by comparing their energy budget with respect to the standard RD energy density. For the universe to transition into standard RD at temperature $T_{\rm RD}$ at the end of a $w$-dominated epoch, there are two main prescriptions: (i) the dominating field $\Phi$ can decay with decay width $\Gamma _{\Phi}$ and the relativistic decay products start dominating the universe as RD. In this case, the Boltzmann equation is
\begin{equation}
\ddot{\rho}_{\Phi} + 3(1+w)H\dot{\rho}_{\Phi}=-\Gamma _{\Phi}\rho_{\Phi}.
\label{BoltzEqPhidecay}
\end{equation}
(ii) If $w>1/3$, then the energy density of the species with EoS $w$ dilutes faster than radiation and therefore radiation takes over naturally. 

In both of these cases, the transition to RD is typically assumed to be instantaneous, but it can be slow depending on the details of model building (e.g., couplings of $\Phi$, model of quintessential inflation, etc.). However, the slow transition has to be treated with varying $w$ rather than a constant EoS, which itself is a complicated analysis. In the next section, PBH formation in a few interesting cases is discussed in detail. The post-inflationary universe can also be dominated by light PBHs, which decay via Hawking radiation to reheat the~universe. 

\section{Primordial Fluctuations}
\label{reason}
Many cosmological scenarios have been proposed that can lead to PBH formation, of which possibly the most popular scenario is when the primordial epoch of inflation leads to scalar fluctuations ($\zeta$: curvature perturbation)\footnote{Here, we will use the uniform density curvature perturbation $\zeta$ and comoving curvature perturbation $\mathcal{R}$ interchangeably, since $-\zeta = \mathcal{R}$ at the superhorizon scales.}, which become frozen soon after they exit the inflationary horizon in the simple case of single field inflation. For multi-field inflationary scenarios, these perturbations grow even in the superhorizon regime until the end of inflation. These perturbations generate classical density fluctuations $\delta ({\bf x},t)=\frac{\rho - \rho _b}{\rho _b}$, with $\rho _b$ being the background energy density, when they re-enter the post-inflationary horizon:
\begin{equation}
	\delta ({\bf x},t)=\frac{2(1+w)}{5+3w}\bigg( \frac{1}{aH}\bigg)^2\bigtriangledown ^2\zeta ({\bf x},t),
	\label{delta1}
\end{equation}
where $w$ is the equation of state of the background at the epoch of re-entry, and $a$ is the scale~factor.

These overdensities grow inside the post-inflationary horizon and the nature of growth is dictated by $w$. The overdense regions of scale $R$ will stop expanding after some time and collapse gravitationally against the pressure if the mass corresponding to $R$ is larger than the Jeans mass. A critical value of the density contrast $\delta _c$ can also be defined, such that the overdensities with $\delta \geq \delta _c$ lead to collapse and form PBH. The value of $\delta _c$ depends on the background and, for the RD epoch, it is $\sim$$0.4$.
If, at the horizon scale, the fluctuations have a Gaussian distribution, then the analysis for collapse is easier, where $\delta _c$ resides at the tail of the distribution. In the case of primordial inflationary fluctuations, a one-to-one relation can be developed between the mass of the PBH produced and the wavenumber $k=2\pi /R$, and this relation crucially depends on $w$. Many examples in the literature are devoted to envisaging inflationary scenarios where the scalar fluctuations grow to large values during inflation such that the power spectrum $\mathcal{P}_{\zeta} (k)$ peaks around a certain wavenumber $k_p \gg k_{\rm CMB}$, which can lead to PBH of mass $M_{k_p}$. If $\mathcal{P}_{\zeta} (k)$ has a broad peak around $k_p$, which is the case in realistic inflationary models, then the PBH mass spectrum is also broad. 
%\begin{center}

\begin{figure}[H]
\includegraphics[width=0.9\textwidth]{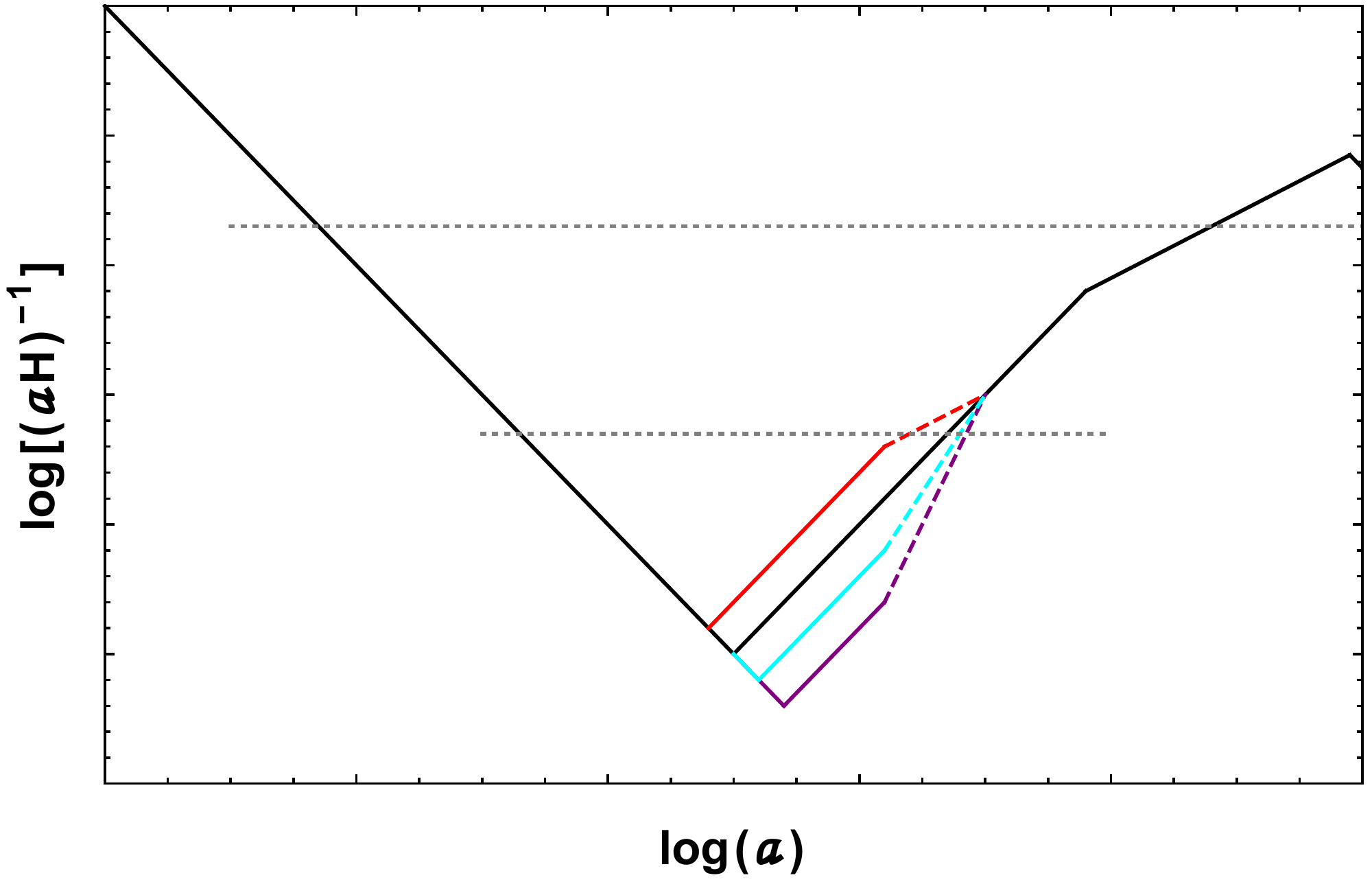}
\caption{Evolution of the horizon scale $(aH)^{-1}$ with the scale factor (both plotted in $\log$). The standard $\Lambda$CDM+Inflation evolution is shown with the solid black line. Red, cyan and purple plots signify alternate evolutions in the presence of a non-standard post-inflationary epoch with $w=0$,$2/3$ and $1$, respectively. Evolutions in the non-standard epochs are shown with dashed lines, whereas standard but shifted evolutions before that are shown with colored solid lines. Dotted grey lines represent cosmological~scales.}
\label{evol}
\end{figure}

%\end{center}
In Figure~\ref{evol}, the evolution of the horizon is shown as a function of the scale factor for the standard case and with the inclusion of a non-standard $w$-dominated epoch after inflation. In this case, instantaneous reheating is assumed for simplicity. The two length scales plotted in dotted grey lines exit the horizon during inflation, larger scale first, and re-enter the post-inflationary horizon, smaller scale first. The smaller scale plotted in Figure~\ref{evol} is such that it enters during $w$-domination. Depending on $w$, the scale factor $a_{\rm hc}$ at the time of horizon crossing of this scale is different, and as a result, $H_{\rm hc}$ depends on $w$. Therefore, the PBH mass $M$ formed due to the collapse of a mode $k$ depends on $w$. This has been discussed in detail in Section~\ref{Mandk}. One interesting outcome in the presence of a non-standard epoch is that the RD evolution before the onset of $w$-domination gets shifted (see Figure~\ref{evol}). Moreover, the evolution in the inflationary epoch decreases or increases for a softer or harder EoS with respect to RD, respectively. This has important implications in terms of inflationary observables in CMB since the duration of inflation affects the scalar spectral index $n_s$ and tensor-to-scalar ratio $r$.

Large primordial fluctuations are necessary for abundant PBH. For example, assuming Gaussian probability distribution for the primordial fluctuations, $\mathcal{P}_{\zeta}\sim 10^{-2}$ is required to reach at least a percent level contribution of PBH into total DM when PBHs are formed in a RD epoch (see the derivations in Section~\ref{PBHformw} and Table~\ref{tablefpbh} in Section~\ref{pbhsecmd}). For single-field models of inflation, if the inflaton slows down enough in its potential, then ultra slow-roll (USR) conditions can be reached. In this case, the slow roll parameters $\epsilon _V\equiv \frac{M_P^2}{2}\bigg(\frac{V_{\phi}}{V}\bigg)^2$ become extremely tiny as compared to its value at CMB, and $\eta _V\equiv M_P^2\frac{V_{\phi \phi}}{V}$ attains a large negative value $\eta _V \leq -6$. Since $\mathcal{P}_{\zeta}\propto 1/\epsilon _V$, to reach from $\mathcal{P}_{\zeta}\sim 10^{-9}$ at CMB scales to $\mathcal{P}_{\zeta}\sim 10^{-2}$ at a smaller scales, $\epsilon _V$ needs to decrease by $\sim$$10^7$ orders in magnitude. This USR mechanism leads to the growth of perturbations and therefore large $\mathcal{P}_{\zeta}$. The modes for which $\mathcal{P}_{\zeta}$ is large are separated from the CMB modes since $\mathcal{P}_{\zeta}$ is constrained by the Planck survey to have an amplitude $\sim$$10^{-9}$ and a red-tilt at CMB scales. For such models, the steepest growth in $\mathcal{P}_{\zeta}(k)$ is $\sim$$k^4$. To reach such an USR condition, many single field inflation scenarios are modelled with a point of inflection~\cite{Garcia-Bellido:2017mdw,Ballesteros:2017fsr,Ballesteros:2020qam,Bhaumik:2019tvl,Germani:2017bcs,Gangopadhyay:2021kmf} or a tiny bump (or dip)~\cite{Mishra:2019pzq} such that the field velocity is negligible for a range of e-folds $\Delta N$. There can also be scenarios, where the non-canonical kinetic energy of the inflaton can lead to a decrease in the speed of sound, leading to interesting results for PBH formation~\cite{Zhai:2022mpi,Kamenshchik:2018sig,Kamenshchik:2021kcw,Gorji:2021isn,Solbi:2021wbo,Solbi:2021rse}. PBHs can also be formed for inflation models arising from scalar-tensor theories~\cite{Yi:2022anu}, with non-minimal derivative coupling~\cite{Heydari:2021gea}, from squeezed initial states~\cite{Ragavendra:2020vud}, etc.

For multi-field models of inflation, the coupling with a secondary field can induce a large $\mathcal{P}_{\zeta}(k)$, since the inflaton velocity now depends on the full multi-field potential. As an example, for the hybrid inflation model~\cite{Clesse:2015wea}, the mild waterfall phase leads to the growth of $\mathcal{P}_{\zeta}(k)$. For multiple fields present during inflation, there are other avenues that can lead to growth in $\mathcal{P}_{\zeta}(k)$, such as non-trivial coupling to gravity, non-canonical coupling of the inflaton and the secondary field~\cite{Braglia:2020eai,Braglia:2020taf} or from a large turning rate in the field space~\cite{Flauger:2009ab,Flauger:2014ana,Gao:2015aba,Fumagalli:2020adf,Fumagalli:2020nvq,Braglia:2020taf,Fumagalli:2021cel,Fumagalli:2021dtd,Bhattacharya:2022fze}, via inducing instabilities in the isocurvature fluctuations that can be transferred to curvature fluctuations \cite{Ashoorioon:2019xqc}, etc. In the case of warm inflation models, the energetic coupling between the inflaton and the thermal bath can lead to enhancement in $\mathcal{P}_{\zeta}(k)$~\cite{Arya:2019wck,Correa:2022ngq}. In the case of PBH formation, necessary large quantum fluctuations can backreact on the long wavelength modes, and therefore the inflationary dynamics can be discussed in terms of stochastic inflation~\cite{Vennin:2020kng,Animali:2022otk,Ando:2020fjm}.

A treatise of inflation models leading to large scalar perturbations and eventually to PBH can also be found in some other interesting reviews in this issue.

Other interesting methods of PBH production include the collapse of cosmic loops, collapse through bubble nucleation, the collapse of Q-balls, and domain walls, etc. (see Section~\ref{intro} for references). The collapse mechanism can also be discussed as a critical phenomena~\cite{Niemeyer_1998,Niemeyer_1999,Musco_2005,Musco_2009} where the mass of the PBH depends on the overdensity via a critical parameter $\xi$, such that
\begin{equation}
M\propto (\delta -\delta _c)^{\xi},
\end{equation}
where $\delta _c$ is the critical overdensity.
%---------------------------------------------------------------------------------------------

\section{Formation of PBH: Analysis}
\label{analysis}
The formation of PBH from large density fluctuations is a probabilistically rare process. This is because the overdensity, defined as $\delta \equiv \frac{\rho - \bar{\rho}}{\rho}$ where $\rho$ and $\bar{\rho}$ are the local and average densities, can be very large only at the tails of the probability distribution. This process is quantified by defining a threshold of PBH formation with the critical value of the overdensity, $\delta _c$, such that only $\delta \geq \delta _c $ can result in collapse into a PBH. It will be clear from the discussions of the current section that the dependence of $\delta _c$ on the background EoS significantly influences PBH abundance.  
It is evident from the discussion in previous sections that many components contribute to the formation of PBH. In this section, the dependence of these components on the background EoS is discussed, and the relevant mathematics is explained. The mechanism for PBH formation in a general $w$-dependent epoch was first discussed in~\cite{Bhattacharya:2019bvk}. While developing the mechanism and presenting the results in the next section, the focus is on non-rotating PBHs, which neither lose any mass due to Hawking radiation nor accrete\footnote{This is a simplified assumption, since light PBHs of mass $\gtrsim$$10^{15}$ gm have a significant mass loss due to radiation, whereas heavier PBHs of near solar mass tend to accrete and merge.}. 

This subsection contains three main parts. In Section~\ref{Mandk}, the general $w$-dependent relation between the PBH mass and wavenumber is formulated. In Section~\ref{PBHformw}, PBH mass spectrum has been developed for a general $w$-dependent epoch. Here, matter-dominated formation is treated separately since there is no pressure to counter the inward gravitational pull during PBH formation in this epoch. In Section~\ref{contri}, various quantities and mechanisms contributing to the PBH mass spectrum are discussed in detail, including the merits and demerits of simple assumptions that are generally used for these quantities.
\subsection{Length Scale and PBH Mass}
\label{Mandk}
If overdensity corresponding to a scale with wavenumber $k$ leads to PBH of mass $M$, then one can find a relation between these two given a particular background epoch. If $H$ is the Hubble parameter at the time of horizon entry of the mode $k$ in the post-inflationary epoch with EoS $w$, then $H\propto a^{-3(1+w)/2}$, and from $k=aH$, one can find
\begin{equation}
k\propto H^{\frac{1+3w}{3(1+w)}}.
\label{kH1}
\end{equation}
The total mass within the horizon of size $H^{-1}$ is $M_{H}=\frac{4\pi H^{-3}\rho}{3}$, and only a  fraction\footnote{Typically assumed to be $\gamma =0.33$, although, it can depend on the epoch of formation~\cite{Koike:1995jm}.} of this mass is collapsed to form PBHs: $M=\gamma M_{H}$. Using the Friedman equation $H^2 = \frac{\rho}{3M_P^2}$, the mass and Hubble parameter are related as
\begin{equation}
M=\frac{4\pi \gamma M_P^2}{H}.
\label{MassHub}
\end{equation}
Then, the dependence of $M$ on $k$ is
\begin{equation}
M\propto \bigg(\frac{k}{4\pi \gamma M_P^2}\bigg)^{\frac{1+3w}{3(1+w)}}.
\label{Mkapprox}
\end{equation}

An exact relation between $M$ and $k$ can also be found. The exact dependence of $k$ and $M(k)$ on the temperature $T$ at formation can be found using the matching relations of the form
\begin{equation}
H(T)=\frac{H(T)}{H(T_{\rm RD})}H(T_{\rm RD})=\bigg(\frac{a(T)}{a(T_{\rm RD})}\bigg)^{-\frac{3(1+w)}{2}}\bigg(\frac{\pi ^2 g_*(T_{\rm RD})}{45M_P^2}\bigg)^{1/2}T_{\rm RD}^2,
\end{equation}
where $\rho (T_{\rm RD})=\rho _{R}(T_{\rm RD})+\rho _w{T_{\rm RD}}=2\rho (T_{\rm RD})=2\frac{\pi ^2}{30}g_*(T_{\rm RD})T_{\rm RD}^4$. Here, $g_*(T)$ and $g_s(T)$ denote the energy and entropy degrees of freedom, respectively. Using the conservation of entropy $g_s(T)a(T)^3T^3$ at every epoch, one can find $k=a(T)H(T)$ to be
\begin{equation}
k=\bigg(\frac{\pi ^2 g_*(T_{\rm RD})}{45M_P^2}\bigg)^{1/2} a_{\rm eq}T_{\rm eq} \bigg(\frac{g_s(T)}{g_s(T_{\rm RD})}\bigg)^{\frac{1+w}{2}}\bigg(\frac{g_s(T_{\rm eq})}{g_s(T)}\bigg)^{\frac{1}{3}}T^{\frac{1+3w}{2}}T_{\rm RD}^{\frac{1-3w}{2}},
\label{relkT}
\end{equation}
where the subscript `eq' corresponds to the time of matter radiation equality in standard cosmology.
This leads to the following expression for $M(k)$:
\begin{equation}
M(k)=4\pi \gamma M_P^2 \bigg(\frac{\pi ^2 g_*(T_{\rm RD})}{45M_P^2}\bigg)^{\frac{1}{1+3w}} \bigg(\frac{g_s(T_{\rm eq})}{g_s(T_{\rm RD})}\bigg)^{\frac{1+w}{1+3w}}(a_{\rm eq}T_{\rm eq})^{\frac{3(1+w)}{1+3w}} T_{\rm RD}^{\frac{1-3w}{1+3w}}k^{-\frac{3(1+w)}{1+3w}}.
\label{Mkexact}
\end{equation}

\begin{figure}[H]
\includegraphics[width=0.9\textwidth]{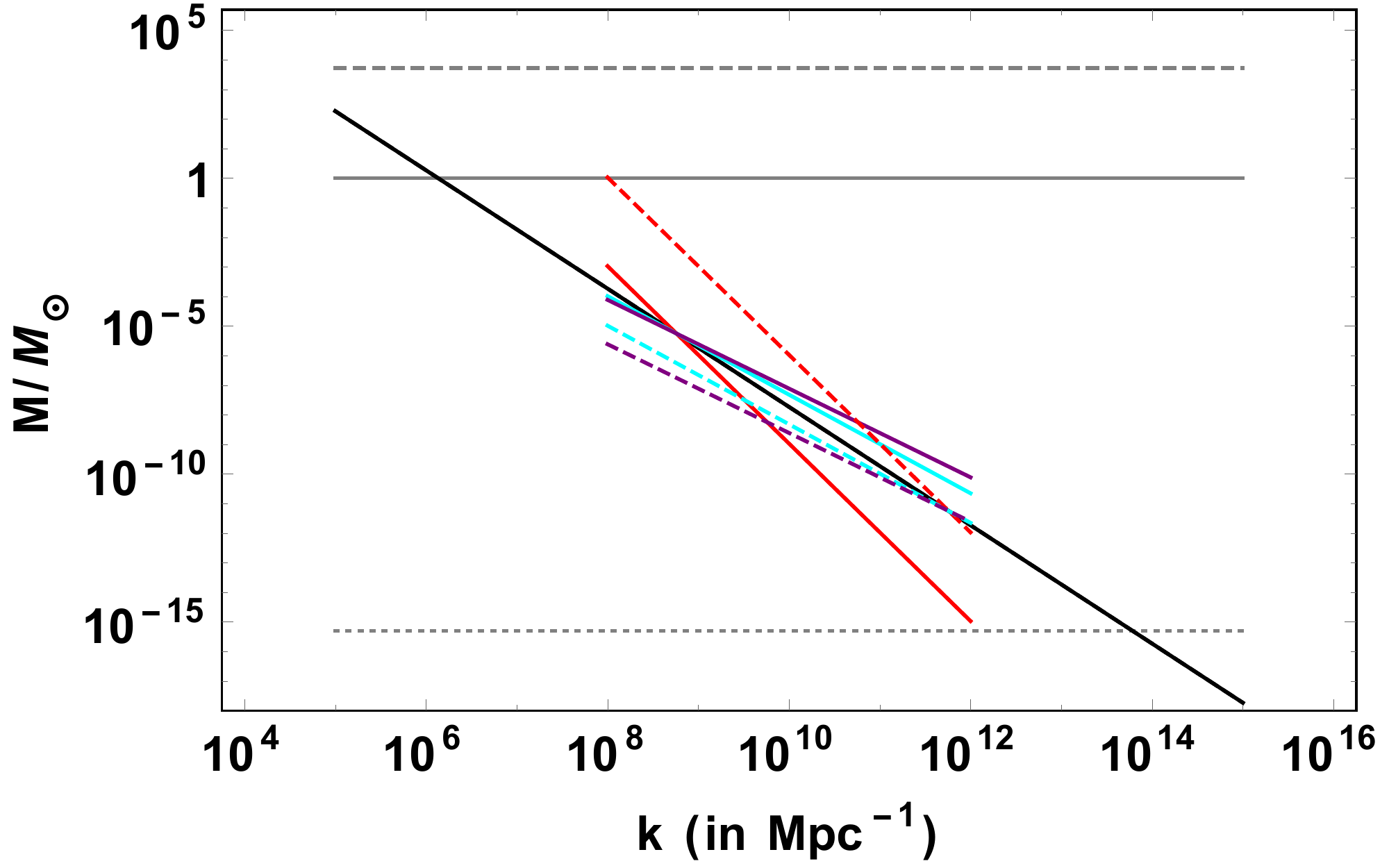}
\caption{PBH mass $M(k)$ with the mode $k$ for different values of $w$ and $T_{\rm RD}$. Red, cyan and blue lines signify $w=0$,$2/3$ and $1$, respectively, for a smaller range of modes larger than $k_{\rm BBN}$, whereas the black line indicates standard RD formation for a larger range in $k$. Solid and dashed lines are for $T_{\rm RD}=100$~GeV and $T_{\rm RD}=10^5$ GeV, respectively. The solar mass, the largest possible PBH mass formed at $T_{\rm BBN}=5$~MeV and $M=10^{15}$ gm are shown with gray solid, dashed and dotted lines, respectively.}
\label{Mvsk}
\end{figure}

This dependence has been elaborated with reference to Figure~\ref{evol} in the previous section. The dependence in Equation~\eqref{Mkexact} can be written in the following convenient form:
\begin{equation}
\frac{M(k)}{M_{\odot}}=4\pi \gamma C(w) \bigg(\frac{T_{\rm RD}}{GeV}\bigg)^{\frac{1-3w}{1+3w}}\bigg(\frac{k}{Mpc^{-1}}\bigg)^{-\frac{3(1+w)}{1+3w}},
\label{Mkexactsol}
\end{equation}
where $C(w)$ is a numerical factor for a particular $w$. Figure~\ref{Mvsk} shows the possible PBH masses given particular values of $w$ and $T_{RD}$. From this figure, we note that, for a particular value of k, the mass of the PBH formed depends crucially on $w$ and $T_{\rm RD}$. The mass range in which PBHs are formed for a particular range in $k$ always decreases with $w$, for a fixed value of $T_{\rm RD}$. 
%\begin{center}
%\end{center}

\subsection{Formation in a $w$-Dominated Epoch}
\label{PBHformw}
If the probability of the gravitational collapse of an overdensity $\delta$ to a PBH is $P(\delta)$, then the probability that PBH of mass $M$ has formed is given by the mass fraction $\beta (M)$. If the formation takes place in a $w$-dominated epoch, then $\beta (M)$ depends on $w$ via the
% related to $w$ via the 
critical overdensity $\delta _c(w)$, since, using Press--Schechter formalism,
\begin{equation}\label{eq:beta}
	\beta(M) = \int_{\delta_c}^{\infty} d\delta ~P(\delta).
\end{equation}
If the density fluctuations have a Gaussian profile, then 
\begin{equation}
P(\delta)= \frac{2}{\sqrt{2\pi}\sigma(M)} \exp\left(-\frac{\delta^2}{\sigma(M)^2}\right),
\label{probdel}
\end{equation}
where $\sigma (M)$ is the variance of the density fluctuation for a scale relating to PBH mass $M$ and can be written in terms of the primordial curvature power spectrum $P_{\zeta}(k)$ as 
%\cad{REFERENCE}
\begin{equation}
	\sigma^2(M)=\frac{4(1+w)^2}{(5+3w)^2}\int \frac{dk}{k}(kR)^4 W^2(k,R)P_{\zeta}(k).\label{sigmaM1}
\end{equation}
Choosing a Gaussian window function $W(k,R)$ smoothens the perturbations on the comoving scale $R$ at formation. Therefore, the approximate relation that can be used is 
\begin{equation}
	\sigma (M) \simeq \frac{2(1+w)}{(5+3w)}\sqrt{P_{\zeta}(k)}.
	\label{sigmaM2}
\end{equation}

The fraction of the background energy density that collapses into forming a PBH is $\frac{\rho _{\rm PBH}}{\rho}\vert_{i}=\gamma \beta (M)$, where the subscript $i$ defines the time of formation of PBH of mass $M$, and $\rho$ is the total energy density of the universe at the time of formation. The fraction of DM in the form of PBH, i.e., PBH abundance, is defined through the PBH mass function $\psi (M)$
as 
\begin{equation}\label{eq:psi}
	\psi (M)=\frac{1}{M}\frac{\Omega_{\rm PBH}(M)}{\Omega _{\rm DM}}\bigg \vert _0\,.
\end{equation}
$\psi (M)$ is related to the fractional energy in the form of PBH at formation. Using the evolution of the energy in PBH after formation until the epoch of matter--radiation equality (denoted with suffix `eq'), the mass function today can be determined, since PBH energy density and the background energy density evolve similarly in a MD epoch\footnote{Here, we neglect the formation of PBHs of mass $M$ via collapse or accretion anytime after the primordial formation. The recent epoch of dark energy domination can be neglected as well since it started dominating at around redshift $z\simeq 1$.}.
\begin{eqnarray}
\psi (M)&=&\frac{1}{M}\frac{\Omega_{\rm PBH}(M)}{\Omega_c}=\frac{1}{M}\frac{\rho_{\rm PBH}(M)}{\rho_c}\bigg\vert_{\rm eq}\nonumber \\
&=&\frac{1}{M}\frac{\rho_{\rm PBH}(M)}{\rho_{\rm rad}}\bigg\vert_{\rm eq}\bigg(\frac{\Omega_m h^2}{\Omega_c h^2}\bigg)\nonumber \\
&=&\frac{1}{M}\frac{\rho_{\rm PBH}(M)}{\rho_{w}}\bigg\vert_{T_{\rm RD}}\bigg(\frac{a(T_{\rm eq})}{a(T_{\rm RD})}\bigg)\bigg(\frac{\Omega_m h^2}{\Omega_c h^2}\bigg)\nonumber \\
&=&\frac{1}{M}\frac{\rho_{\rm PBH}(M)}{\rho_{w}}\bigg\vert_{T}\bigg(\frac{a(T_{\rm RD})}{a(T)}\bigg)^{3w}\bigg(\frac{a(T_{\rm eq})}{a(T_{\rm RD})}\bigg)\bigg(\frac{\Omega_m h^2}{\Omega_c h^2}\bigg)\nonumber \\
&=& \frac{\gamma \beta(M)}{M}\bigg(\frac{g_s(T_{\rm RD})}{g_s(T)}\bigg)^{-w}\bigg(\frac{g_s(T_{\rm eq})}{g_s(T_{\rm RD})}\bigg)^{-1/3}\bigg(\frac{T}{T_{\rm RD}}\bigg)^{3w}\bigg(\frac{T_{\rm RD}}{T_{\rm eq}}\bigg)\bigg(\frac{\Omega_m h^2}{\Omega_c h^2}\bigg).
\label{psiM0}
\end{eqnarray}
Here, in the second line, we have used the condition that matter and radiation energy density are equal at $T_{\rm eq}$. Similarly, in the third line, we have used the equality of the radiation energy density and the energy density of the species with EoS $w$ at $T_{\rm RD}$. In the last line, we have used the conservation of entropy. Using the relations between $k$ and $T$ in Equation~\eqref{relkT} and the expression for $M(k)$ in Equation~\eqref{Mkexact}, the mass function can be written in terms of the PBH mass $M$ as
%\begin{adjustwidth}{-\extralength}{0cm}
%\centering %% If there is a figure in wide page, please release command \centering
\begin{equation}
\psi (M)=\frac{\gamma}{T_{\rm eq}}(4\pi \gamma M_P^2)^{\frac{2w}{1+w}}\bigg(\frac{g_s(T_{\rm RD})}{g_s(T_{\rm eq})}\bigg)^{1/3}\bigg(\frac{\pi ^2 g_*(T_{\rm RD})}{45M_P^2}\bigg)^{-\frac{w}{1+w}}\bigg(\frac{\Omega_m h^2}{\Omega_c h^2}\bigg)T_{\rm RD}^{\frac{1-3w}{1+w}}\beta (M)M^{-\frac{1+3w}{1+w}}.
\label{MpsiMw}
\end{equation}
%\end{adjustwidth}

The total contribution of PBH to the DM abundance can be evaluated now as
\begin{equation}
f_{\rm PBH}=\int dM \psi (M).
\label{ftot}
\end{equation}
Thus, the quantity $M \psi (M)d\ln M$ can also be viewed as the fractional PBH abundance in the logarithmic mass range from $\ln M$ to $\ln (M+\delta M)$.
%------------------------------------------------------------------------------------------------------------
\subsection{Formation in a Matter Dominated Epoch}
\label{formmd}
The difference between PBH formation in a MD epoch is different than what is discussed above since the collapse dynamics are different in the absence of pressure. For an overdense region collapsing in a pressureless background, the gravitational pull aiding in the collapse is not contested by pressure. As a result, the sphericity of an initially spherical overdense region gets affected, as is discussed in detail in~\cite{Harada:2016mhb}, which uses Zel'dovich approximation, Thorne's hoop conjecture, and Doroshkevich's probability distribution to compute the mass fraction of PBH in a MD epoch. In this case, for perturbations of order $\sigma \leq 0.01$, the mass fraction
was found to be
\begin{equation}
\beta _{\rm MD}(M)\simeq 0.056\sigma (M)^5.
\label{betaMD}
\end{equation}

The mass function can then be found by putting $w=0$ in Equation~\eqref{psiM0}.
Due to the complete absence of pressure, PBHs formed in a MD epoch can have large spins~\cite{Kuhnel:2019zbc}, the accretion dynamics~\cite{DeLuca:2021pls} and clustering~\cite{Matsubara:2019qzv} can be different, and the ellipticity can also affect the formation process~\cite{Yoo:2020lmg,Kokubu:2018fxy}.
%, with $\epsilon =1$.
The growth of the overdensities after horizon re-entry depends crucially on the EoS $w$. The formation in a MD epoch has to be discussed explicitly since density perturbations grow linearly in MD so that $\delta \sim \sqrt{\langle \sigma \rangle ^2} \sim a$, where $a$ is the scale factor. Here, $\sigma ^2 $ is the variance in $\delta$-distribution. It becomes nonlinear when $\delta \sim \mathcal{O}(1)$. If $\sigma$ is defined in the linear regime at the time of horizon entry of the modes, then the scale factor $a_m$ at the time $t_m$ of maximum expansion is given by $\sigma (a_m)/a_{\rm hc}$, where $a_{\rm hc}$ is the scale factor at the time $t_{\rm hc}$ of horizon entry~\cite{Carr:2017edp,Harada:2017fjm,Nakama:2018utx,Khlopov:1980mg,Polnarev:1985}.
The time of collapse $t_c$ is very close to $t_m$ and, therefore, the scale factor at $t_c$ is $a_c \simeq a_m$. 
Thus, $t_c/t_{\rm hc} = (a_c/a_{\rm hc})^{3/2} = \sigma ^{-3/2}$. 
The Hubble parameters at horizon entry $H_{\rm hc}$ and at the time of collapse $H_c$ are therefore related as $H_{\rm hc}/H_c = \sigma^{-3/2}$. Thus, the PBH that is formed from the mode that enters the horizon at $t_{\rm hc}$ has a mass
\begin{equation}
M=\frac{4\pi \gamma M_P^2}{H_c}\sigma ^{3/2}.
\label{MassHubEMD}
\end{equation}

In the EMD epoch, the PBH mass formed as a result of collapse (when $\sigma$ becomes nonlinear)
at time $t_c$ can be estimated using Equation~\eqref{MassHubEMD}. In comparison, note that the growth of perturbations is logarithmic in a RD epoch, and therefore, the mass of PBHs formed in RD can be estimated by Equation~\eqref{MassHub}.

The PBH mass function in EMD is limited within two mass scales, $M_{\rm max}$ and $M_{\rm min}$, corresponding to the largest and the smallest scales, respectively, that became nonlinear during EMD. $M_{\rm max}$ corresponds to the mode $H^{-1}_{\rm max}$ that entered the horizon at some point before reheating and collapsed at the time of reheating. Therefore, following the arguments in the previous paragraph, $M_{\rm max}$ is given by~\cite{Carr:2017edp}
\begin{equation}
M_{\rm max}=\frac{4\pi \gamma M_P^2}{H_{\rm max}}=\frac{4\pi \gamma M_P^2}{H_{\rm RD}}\sigma ^{3/2}=M_{\rm RD}\sigma ^{3/2},
\label{MmaxHubEMD}
\end{equation}
where $\sigma$ can be found using Equation~\eqref{sigmaM2}, once the primordial power spectrum $\mathcal{P}_{\zeta}(k)$ is specified. However, since $\sigma <1$ always, $M_{\rm max}<M_{\rm RD}$.

Since the growth of perturbations is already accounted for via $\beta (M)$ for MD, $\gamma =1$ while calculating the energy density fraction collapsing into PBHs at the time of formation. For $w\neq 0$, $\gamma$ is a $\mathcal{O}(1)$ parameter~\cite{1978SvA....22..129N}.
PBHs are formed more efficiently in a MD epoch than in a RD epoch due to the power law dependence of the mass fraction $\beta (M)$ on $\sigma (M)$ rather than an exponential dependence, which can be explored for different scenarios of matter or near-dust dominated epochs~\cite{Ballesteros:2019hus,Bhattacharya:2020lhc,Domenech:2020ers,Das:2021wad,Bhattacharya:2021wnk,Choi:2022btl}.
\subsection{Understanding the Contributions}
\label{contri}
In the last section, different quantities have been introduced which contribute to the PBH abundance in DM. Some of these quantities are very relevant from a phenomenological point of view. Some of these quantities are assumed to have simple forms, which is easier to work with when one studies specific scenarios of PBH formation; however, there can be well-motivated scenarios where these assumptions are violated. In this section, a few such quantities are discussed with reference to the validity of their values or forms and their impact on the PBH abundance.

\subsubsection{Critical Overdensity $ \delta _c$} 
Throughout many decades, the effort to compute the critical overdensity (also termed as the density threshold for PBH formation) has been in progress. In 1974 and 1975, Carr and Hawking~\cite{Carr:1974nx,Carr:1975qj} used the Jeans instability criterion in Newtonian gravity to deduce $\delta _c \sim c_s^2$, where $c_s$ is the sound speed, and for a static fluid, $c_s^2 = w$. After that, many attempts have been made with numerical hydrodynamic solutions and more in~\cite{1978SvA....22..129N,1979ApJ...232..670B,1980SvA....24..147N,Niemeyer_1998,shibata99,Hawke:2002rf}. In~\cite{Niemeyer_1998}, the lengthscale of the perturbation was measured with a Gaussian shaped profile for $\delta$, whereas~Ref. \cite{shibata99} measured the local peak of the curvature profile. These two analyses used different approaches and assumptions about the decaying mode of perturbations and reached different conclusions for $\delta _c$ in a RD epoch. In~\cite{Green2004}, $\delta _c$ was measured using the linear relation between curvature and energy density profile. While~Refs. \cite{shibata99,Green2004} measured the local value of $\delta _c$, Refs. \cite{Niemeyer_1998,Musco_2005} measured the average $\delta _c$. In 2013, Harada et al.~\cite{Harada:2013epa} deduced $\delta _c$ analytically using a three-zone model for the overdensity profile by imposing the requirement that the time taken by the pressure sound wave to cross the scale of the overdense region is larger than the time of onset of the gravitational collapse. This work resulted in the following $w$-dependent expression for $\delta _c$ in the comoving gauge, which is used in this review:
\begin{equation}
\delta _c=\frac{3(1+w)}{(5+3w)}\sin ^2 \bigg(\frac{\pi \sqrt{w}}{(1+3w)}\bigg). \label{deltac_w}
\end{equation}

However, the critical value $\delta _c$ also depends crucially on the shape of the density profile, which can be parameterised as
\begin{equation}
\alpha = -\frac{r_m^2\mathcal{C}''(r_m ,t)}{4\mathcal{C}(r_m ,t)},
\label{shapealpha}
\end{equation}
%Primes denote derivs wrt $r$
where $\mathcal{C}(r,t)=\frac{2\delta M (r,t)}{R(r,t)}$ is the compaction function defined as the ratio of the mass excess over the physical radius, and $R=a(t)r$ is the aerial radius of the overdense region. Primes denote derivatives with respect to the position $r$, and $r_m$ is the position where $\mathcal{C}(r,t)$ is maximised. $\alpha\gg1 (\ll 1)$ signifies a broad (narrow) peak. The form of $\delta _c$ in Equation~\eqref{deltac_w}, which does not account for the shape, is more precise for $\alpha \rightarrow 0$, since broader $\delta$ profiles may `bounce back' and disfavor the collapse. 

$\delta _c$ can also be calculated from the compaction function, focussing on the local values of $\delta (r)$ with radius $r$ inside the spherical overdense region. This process thus takes into account the shape of the density profile. Although the calculation of $\delta _c$ by comparing the pressure and gravitational pull using the three-zone model and therefore Equation~\eqref{deltac_w} is very popular, and used in this review, using the compaction function provides more insight into the shape of the peak profile and in general is more useful in scenarios that include nonlinearities and non-Gaussianities.  
%Equation~\eqref{eq:beta} in the previous subsection has been formulated using PS mehcanism. 

In this formalism, one focuses on the peak profile of either the metric perturbation $\zeta (\hat{r})$, or the curvature perturbation $K(r)$~\cite{Musco:2018rwt,Kalaja:2019uju,Germani:2018jgr}. In terms of $K(r)$, the perturbed metric~is:
\begin{equation}
ds^2=-dt^2+a^2(t)\bigg(\frac{dr^2}{1-K(r)r^2}+r^2d\Omega ^2\bigg).
\end{equation}
%The peak density contrast is related to the curvature perturbation as (Equation(3.7) in~\cite{Kalaja:2019uju}):
%\begin{equation}
%\delta _{\rm peak}(\hat{r})=f(w)\frac{2}{3}\bigg(\frac{1}{aH}\bigg)^2e^{2\zeta _{\rm peak}(\hat{r})}\bigg[\nabla ^2\zeta _{\rm peak}(\hat{r})-\frac{1}{2}\nabla \zeta _{\rm peak}(\hat{r}). \nabla \zeta _{\rm peak}(\hat{r})\bigg],\label{fulldeltazeta}
%\end{equation} 
%where $f(w)=\frac{3(1+w)}{(5+3w)}$. Here, a trough in $\zeta (\hat{r})$ points to a peak in $\delta (\hat{r})$. The linear relation in Equation~\eqref{delta_zeta1} at horizon crossing is recovered when both the curvature perturbation and its gradient are small ($\zeta _{\rm peak}\ll 1$ and $\nabla \zeta _{\rm peak} \ll 1$).

In addition, in terms of the metric perturbation $\zeta (\hat{r})$, it is
\begin{equation}
ds^2=-dt^2+a^2(t)e^{2\zeta (\hat{r})}\bigg(d\hat{r}^2+\hat{r}^2d\Omega ^2\bigg).
\end{equation}
The coordinate transformation between $\zeta (\hat{r})$ and $K(r)$ dictates
\begin{eqnarray}
r&=\hat{r}e^{\zeta} ({\hat{r})}\nonumber \\
\frac{dr^2}{\sqrt{1-K(r)r^2}} &= e^{\zeta} ({\hat{r})}d\hat{r}.
\label{krzetar1}
\end{eqnarray}

From the first expression in Equation~\eqref{krzetar1}, the differential relation between $r$ and $\hat{r}$ is obtained to be
\begin{equation}
\frac{dr}{d\hat{r}} = e^{\zeta  (\hat{r})} (1+\hat{r}\zeta '(\hat{r})).
\end{equation}

Thus,  $\zeta (\hat{r})$ and $K(r)$ are related as
\begin{equation}
K(r)r^2 = -\hat{r}\zeta '(\hat{r}) \bigg[2+\hat{r}\zeta '(\hat{r})\bigg].
\end{equation}

The averaged density contrast, which is a more relevant quantity of interest in case of an extended peak profile of $K(r)$, can be written (at horizon crossing) as:
\begin{equation}
\tilde{\delta } (r)=f(w)K(r) r^2, \label{tildel1}
\end{equation}
where $f(w)=\frac{3(1+w)}{(5+3w)}$, and $r$ is the radius of the spherical comoving volume on which it has been averaged. The coordinate origin is at the location of the peak. PBH formation criteria are expressed in terms of the compaction function $\mathcal{C}(r,t)$.
Now, for a particular peak profile of $K(r)$ or $\zeta (\hat{r})$, there are two scales of importance: the scale $r_0$ where the local density contrast crosses zero and the scale $r_m$ where the compaction function reaches the maximum value. Thus, $\tilde{\delta }_0 (r)=f(w)K(r_0) r_0^2$ and $\tilde{\delta }_m (r)=f(w)K(r_m) r_m^2$. The $\delta$ considered in a PS formalism is equivalent to $\tilde{\delta }_0$, but $\tilde{\delta }_m$ and $\tilde{\delta }_0$ are different in general. 

For a particular peak profile for curvature, one can determine $r_0$, $r_m$, $\tilde{\delta }_0$ and $\tilde{\delta }_m$ in terms of the profile parameters. Then, knowing the critical value of $\tilde{\delta }_0$, we can find the critical value for $\tilde{\delta }_m$. The ratio $\frac{\tilde{\delta }_m^c}{\tilde{\delta }_0^c}=\frac{K(r_m)r_m^2}{K(r_0)r_0^2}$ depends on the shape of the curvature profile. $r_m$ is determined by maximising the compaction function (defined after Equation~\eqref{shapealpha}), and $r_0$ is determined from the zero-crossing of the density profile given by 
\begin{equation}
\delta = \bigg(\frac{1}{aH}\bigg)^2f(w)\bigg[K(r)+\frac{r}{3}K'(r)\bigg].
\end{equation} 

For example, for a Gaussian curvature profile:
\begin{equation}
K(r)=\mathcal{A}e^{-\frac{r^2}{2\Delta ^2}},\label{prof1}
\end{equation}
$K(r_m)=\mathcal{A}/e$ at $r_m^2=2\Delta ^2$ and $K(r_0)=\mathcal{A}/e^{3/2}$ at $r_0^2=3\Delta ^2$.
Hence, the numerical formula for $\tilde{\delta }_0^c$ in Equation~\eqref{deltac_w} gives
\begin{equation}
\tilde{\delta }_m^c=\frac{2e^{1/2}}{3}\tilde{\delta }_0^c=\frac{2e^{1/2}}{3} f(w)\sin ^2\bigg(\frac{\pi \sqrt{w}}{1+3w}\bigg).\label{deltacm1}
\end{equation}
The RD values are $\tilde{\delta }_0^{c,{\rm RD}}=0.414$ and $\tilde{\delta }_m^{c,{\rm RD}}=0.455$, whereas the $w=1$ epoch has $\tilde{\delta }_0^{c,w=1}=0.375$ and $\tilde{\delta }_m^{c,w=1}=0.412$.

The exact value of $\delta _c$ is impacted by nonlinearities~\cite{Kawasaki_2019,Young:2019yug,Germani_2020,Young_2020} and \linebreak non-Gaussianities~\cite{Young_2013,Young_2016,Franciolini_2018,Luca_2019,Yoo_2019,Kehagias_2019}. A nonlinear relation between the primordial fluctuations $\zeta$  and the overdensity $\delta$ 
\begin{equation}
\delta ({\bf x},t)=-\frac{2(1+w)}{(5+3w)}\frac{1}{a^2H^2}e^{-2\zeta ({\bf x})}\bigg(\triangledown ^2\zeta ({\bf x}) +\frac{1}{2}\partial _i\zeta({\bf x})\partial ^i \zeta({\bf x})\bigg)
\label{nonlindelzeta}
\end{equation}
can be crucial since large PBH abundance requires very large values of $\zeta$. This nonlinear relation, when taken into consideration, can lead to a non-Gaussian $P(\delta)$ even if the primordial fluctuations were Gaussian. In well-known attempts to include such
nonlinearities using peak theory or threshold statistics, $\delta _c$ is shown to have a few percent difference than its value when the linear relation is used~\cite{Kehagias_2019}, and the PBH abundance is found to be extremely sensitive to the nonlinear effect. In~\cite{Luca_2019}, it was found that, after including nonlinearities, $\mathcal{O}$(2--3) increase in the initial $\mathcal{P}_{\zeta }$ is required to produce the same PBH abundance as when the linear relationship is used. In~\cite{Germani_2020}, nonlinear statistics relevant to finding PBH abundance have been developed using $\mathcal{C}(r,t)$ as the main statistical variable. 

Given the dependence of $\delta _c$ on the shape of the overdensity profile, nonlinearities and non-Gaussianities, peak theory (PT) calculation of the abundance is majorly used in the literature to account for such non-trivialities. However, in this review, Press--Schechter (PS) theory is used to simplify the calculations. A comparison of PS and PT has been discussed in Section~\ref{secbetaM}. Other methods to compute PBH abundance focussing on the density profile have been discussed in~\cite{Kuhnel:2021yic,He:2019cdb,Suyama:2019npc}. In~\cite{Kuhnel:2021yic}, extreme value theory is used since large values of energy density are reached, which lead to a narrower mass profile and peak at a larger mass as compared to other methods using Gaussian profile, although the total abundance is boosted.
%Historically, the first attempt to compute the PBH formation threshold was done by B. Carr and S. Hawking between 1974 and 1975 [31, 33] where by using a
%Newtonian Jeans instability criterion they were led to the conclusion that δ c ∼ w. Afterwards, δ c was studied through numerical hydrodynamic simulations by the pioneering works of Nadezhin, Novikov & Polnarev in 1978 [34], Bicknell & Henriksen in 1979 [35] and Novikov & Polnarev in 1980 [36] and later after a pause of 20 years by high sophisticated numerical codes by Niemeyer & Jedmazik [37] and Shibata & Sasaki [38]. Within the last decade, there have been a huge progress regarding the determination of the PBH formation threshold both at the analytic as well as at the numerical level. In particular, at the analytic level, T.Harada, C-M. Yoo & K. Kohri (HYK) in 2013 [1] refined the δ c value obtained by Carr in 1975 by confronting the gravitational force which pushes the fluid matter of the collapsing overdensity inwards and enhances as such the gravitational collapse with the pressure gradient force
%which in general pushes the fluid outwards, thus disfavoring the collapsing process. At the end, by comparing the time at which the pressure sound wave crosses the overdensity collapsing to a PBH with the onset time of the gravitational collapse they found that the expression for δ c in the comoving gauge reads as

\subsubsection{Density Distribution $P(\delta)$} 
In realistic models of inflation where the fluctuations are large enough to lead to post-inflationary collapse and PBH, such as those discussed in Section~\ref{reason}, the inflationary dynamics are usually complicated. The same mechanism that leads to the growth of perturbations may also contribute to large non-Gaussianities~\cite{Young_2013,Young_2016,Franciolini_2018,Luca_2019,Yoo_2019,Kehagias_2019,Young:2022phe}. Therefore, the viability of a Gaussian $P(\delta)$ that leads to the simple form of for PBH mass fraction in Equation~\eqref{betaMGauss} needs to be checked when one starts from a specific model of inflation:
\begin{equation}
\beta (M)=\mathrm{erfc}\left(\frac{\delta_c}{\sqrt{2}\sigma(M)}\right)
\label{betaMGauss}
\end{equation}

In~\cite{Young_2016,Franciolini_2018}, primordial non-Gaussianities were included to find that the PBH abundance depends very sensitively on the primordial non-Gaussianities, and therefore primordial non-Gaussianities on small scales can have constraints from constraints on the PBH abundance in certain cases. 
It is to be noted here that $\delta _c$, typically being very large ($\lesssim$$\mathcal{O}(1)$), resides at the tail of $P(\delta)$. Therefore, the fluctuations with $\delta >\delta _c$ are rare, albeit present, even in models of inflation with slow-roll maintained throughout the epoch. In such a case, there will be a very small, but nonzero probability of collapse; however, it leads to a very tiny mass fraction $\beta (M)$ and therefore negligible PBH abundance. 

%\subsubsection{$ \sigma ^2(M)$}
While starting from a model of inflation, $\mathcal{P}_{\zeta}(k)$ typically has a certain width, which does not lead to a monochromatic mass function for PBH. In the simplest scenarios, $\mathcal{P}_{\zeta}(k)$ with a peak at $k=k_p$ can be approximated in a Gaussian form near the peak as
\begin{equation}
\mathcal{P}_{\zeta}(k) = P_0 \exp [-\frac{(\log (k/k_p))^2}{2\sigma _{\zeta}^2}].
\label{PzetaGauss}
\end{equation}

Therefore, in this case, the actual $\sigma (M)$ in Equation~\eqref{sigmaM1} can be significantly different from the approximation in Equation~\eqref{sigmaM2}. The variance of the window function $W(k,R)$ should also be chosen judiciously, depending on $\sigma _{\zeta}^2$. In models of multi-field inflation, a resonant oscillation in $\mathcal{P}_{\zeta}(k)$ around the peak is a common feature that can originate from turns in the field space manifold. In this case, smoothening with the window functions needs to be conducted with caution.  

\subsubsection{Various Methods to Calculate $ \beta (M)$}
\label{secbetaM}
There are several methods to calculate the mass fraction of PBH formation given $\delta _c$, of which Press--Schechter formalism (PS) and Peak theory (PT)~\cite{1986ApJ...304...15B,Yoo:2020dkz,Ferrante:2022mui}
 have gained the most popularity. Whereas the PS method uses the average value of $\delta$ in an overdense region to compare with the critical overdensity to evaluate the PBH abundance, PT focuses on the local distribution of the overdensities, and therefore takes a probabilistic approach to count the number of overdensity peaks. Equation~\eqref{eq:beta} in the previous subsection has been formulated using the PS mechanism. 

Naively, the curvature perturbation $\zeta$ is expected to be the relevant variable for the Gaussian distribution, which is one of the basic assumptions in PS formalism. However, while taking into account the local distribution of fluctuations, the absolute value of $\zeta$ is not relevant, and this brings into question the necessity of a proper statistical variable. In PT, the statistical approach is derived in terms of a much more reliable variable $\nu \equiv \frac{\delta}{\delta _{\rm rms}}$, where $\delta _{\rm rms}$ is the root mean squared value of the density fluctuations. In the simplest scenario, $\delta$ is assumed to be a Gaussian random variable, although non-Gaussianities can be incorporated in the analysis, as discussed in~\cite{Ferrante:2022mui}.
The differential number density $\mathcal{N}_{\rm pk}(\nu)d\nu$ of overdense peaks for Gaussian $\nu$ can be written as
\begin{equation}
\mathcal{N}_{\rm pk}(\nu)d\nu = \frac{1}{(2\pi)^2R_*^3}e^{-\nu ^2/2}G(\tilde{\gamma},\tilde{\gamma}\nu).
\end{equation}
Here, the function $G(\tilde{\gamma},x_*)$ can be written in terms of a fitting formula for large $\nu$ as
\begin{equation}
G(\tilde{\gamma},x_*)=\frac{x_*^3-3\tilde{\gamma}^2x_*+(B(\tilde{\gamma})x_*^2+C_1(\tilde{\gamma})) \exp (-A(\tilde{\gamma})x_*^2)}{1+C_2(\tilde{\gamma})\exp(-C_3(\tilde{\gamma})x_*)},
\end{equation}
where $A(\tilde{\gamma}), B(\tilde{\gamma}), C_{1,2,3}(\tilde{\gamma})$ are specific numerical functions of $\tilde{\gamma}$. $\tilde{\gamma}$ and $R_*$ are spectral parameters which are related to various moments of the power spectrum of density~perturbations
\begin{eqnarray}
\tilde{\gamma} &\equiv & \frac{\sigma _1^2}{\sigma _2 \sigma _0} \nonumber \\
R_* &\equiv & \sqrt(3)\frac{\sigma _1}{\sigma _2}\nonumber {\rm where}\\
\sigma _j^2 &\equiv & \int \frac{k^2dk}{2\pi ^2}k^{2j}\mathcal{P}_{\delta}(k)W_{\delta}^2(kR).
\end{eqnarray}

From the differential number density of peaks, the number density of the overdensity peaks can be written as
\begin{equation}
n_{\rm pk} (\nu _c) = \int _{\nu _c}^{\infty} \mathcal{N}_{\rm pk}(\nu)d\nu .
\end{equation}
For high peaks, it can be calculated as
\begin{equation}
n_{\rm pk} (\nu _c) = \frac{1}{(2\pi ) ^2}\bigg(\frac{\sigma _1 ^2}{3\sigma _0 ^2}\bigg) ^{3/2} (\nu _c ^2 -1)e^{-\nu _c^2 /2}.
\end{equation}
Therefore, the fraction of the PBHs to the total density at the time of formation is given~by 
\begin{equation}
\beta _{\rm PT}(M) d\log M = \frac{M n_{\rm pk}(\nu _c)}{\rho a^3}d\log M.
\end{equation}

It is to be noted here that $n_{\rm pk}(\nu _c)$ depends on the wavenumber $k$ via the moments $\sigma _j^2$, and therefore, given the density power spectrum $\mathcal{P}_{\delta}(k)$, the mass function $\beta _{\rm PT}(M)$ derived using PT depends on $M$ in a complicated manner in general.

\subsubsection{Constant $w$}
The analysis detailed in Section~\ref{PBHformw} as well as most of the literature discussing PBH formation in a non-standard epoch consider the EoS $w$ to be constant during that epoch. \mbox{Section~\ref{PBHformw}} considers the simplest case where the universe also transitions from $w$-domination to RD instantaneously. For example, in models where a heavy field dominates the energy density with $w=0$, the decay of the field to relativistic particles (reheating) is considered to be instantaneous for simplicity. However, in practice, for almost all of the non-standard scenarios, $w$ is not constant. Even if it can be assumed to be constant for most of the non-standard evolution, the transition to RD usually happens over a certain duration of time, which can be modelled by interpolating $w$ between the non-standard value and $1/3$ for RD. However, the dynamics of PBH formation become complicated for a dynamic $w$. PBH formation during slow reheating after inflation, where the EoS slowly transitions from 0 to $1/3$, has been explored in~\cite{Carr:2018nkm}. They found that the mechanism gradually changes from the MD to RD case for $\sigma < \sigma _c =0.005$, below which the mechanism is affected even before the end of reheating. Using $T_{\rm reh}=4$ MeV, they have found that the heaviest PBH that can be produced in the critical case with $\sigma _c =0.005$ is $\sim$$100 \ms$. PBH formation in a (p)reheating epoch is also discussed in~\cite{Padilla:2021zgm,Carrion:2021yeh,Auclair:2020csm,Martin:2019nuw}.
Recently,~Ref. \cite{papanikolaou:2022cvo} showed that, in this case, it is necessary to solve for the critical overdensity $\delta _c$ numerically, with piecewise solutions in terms of the conformal time $\tau$. 
 
%\subsection*{PBH mass shift in MD}

%When comparing the improvement in PBH formation in a $w$-dominated epoch, an useful quantity is the gain in mass function as compared to PBH fomation in a RD epoch. 
%\begin{equation}
%g_w(M)=\frac{\psi (M)}{\psi (M)\vert _{w=1/3}}.
%\end{equation}

\section{Results for Specific Cases}
\label{PBHspecific}
Different possible scenarios where a non-standard post-inflationary epoch can exist have been discussed in Section~\ref{nonst}. The EoS in such an epoch depends on the dominant component of energy density. There are certain well-motivated scenarios where the non-standard epoch is relevant for boosting PBH production. In this section, results for the PBH mass fraction for some specific interesting cases of non-standard post-inflationary epochs are discussed. 

We demonstrate our results using the following two forms of the primordial power spectrum near the peak at $k=k_p$ that are widely used to model the inflationary power spectra without starting from a particular model. While presenting the results, we use $\gamma =0.33$ and $T_{\rm RD} = 100$ GeV.
\subsection{Gaussian Power Spectrum}
In many models of smooth waterfall hybrid inflation~\cite{Clesse:2015wea} and several inflection point models of inflation~\cite{Germani:2017bcs}, the potential features a plateau for a few e-folds before the end of inflation. This plateau regime of the potential can lead to a peak in the curvature power spectrum, which, at the simplest approach, can be written as a Gaussian power spectrum (GPS) of the following form:
\begin{equation}
\mathcal{P}_{\zeta}(k) = P_0 \exp [-\frac{(\log (k/k_p))^2}{2\sigma _{\zeta}^2}].
\label{powG}
\end{equation}

In order to demonstrate the results, $\sigma _{\zeta}=1$ has been used.
\subsection{Broken Power Law Power Spectrum}
In various scenarios of the early universe where PBH is produced from domain walls or vacuum bubbles~\cite{Deng:2016vzb,Deng:2017uwc}, the relevant primordial curvature power spectrum has a broken power law (BPS) form such as:
\begin{equation}
P_{\zeta}(k)=\left\{ \begin{array}{l l} P_0\bigg(\frac{k}{k_p}\bigg)^m & \ \quad k<k_p,\\
P_0\bigg(\frac{k}{k_p}\bigg)^{-n} & \ \quad k\geq k_p \end{array}\right.
\label{powB}
\end{equation}

In order to demonstrate the results, $m=3$ and $n=0.5$ have been used. In Equations~\eqref{powG} and~\eqref{powB}, only the form of the power spectra near the peak are represented. Whenever necessary, the CMB consistent part $A_s (k/k_*)^{n_s-1}$ needs to be added to both of them to obtain the full power spectra.
\subsection{Kinetic Energy Dominated Epoch}
\label{seckdom}
In a model of quintessential inflation, where the scalar field $\phi$ performs the role of inflaton in the early universe and of dark energy in the late universe with different forms of the potential, the inflaton needs to survive at the end of inflation and non-trivial reheating processes need to be implemented. The field $\phi$ needs to travel between the two forms of the potential at early and late times, which can lead to a fast roll of $\phi$ in the intermediate regime. This gives rise to a large kinetic energy of $\phi$, which can come to dominate the universe for some time. During the epoch of such kinetic energy domination (KD), the pressure $p\simeq \rho$, such that the EoS of the epoch is $w\simeq 1$.

In~\cite{Bhattacharya:2019bvk}, the mechanism of PBH formation in a non-standard post-inflationary epoch was applied to $w=1$ for three different types of power spectra to show that, in order to achieve the same PBH abundance, formation in a KD epoch requires less peak amplitude of the primordial power spectrum.
If PBHs are formed due to the overdensities entering in this $w=1$ epoch, then the resulting modification in $\beta (M)$ and $\psi (M)$, as compared to RD formation of PBH, can be evaluated using Equations~\eqref{betaMGauss} and~\eqref{psiM0}.

%In the following, we assume that the KD epoch ends at $T_{\rm RD}=100$ GeV. For both KD and RD epochs, $P_0=0.02$ has been used.

\begin{figure}[H]
\begin{center}
\includegraphics[width=0.5\textwidth]{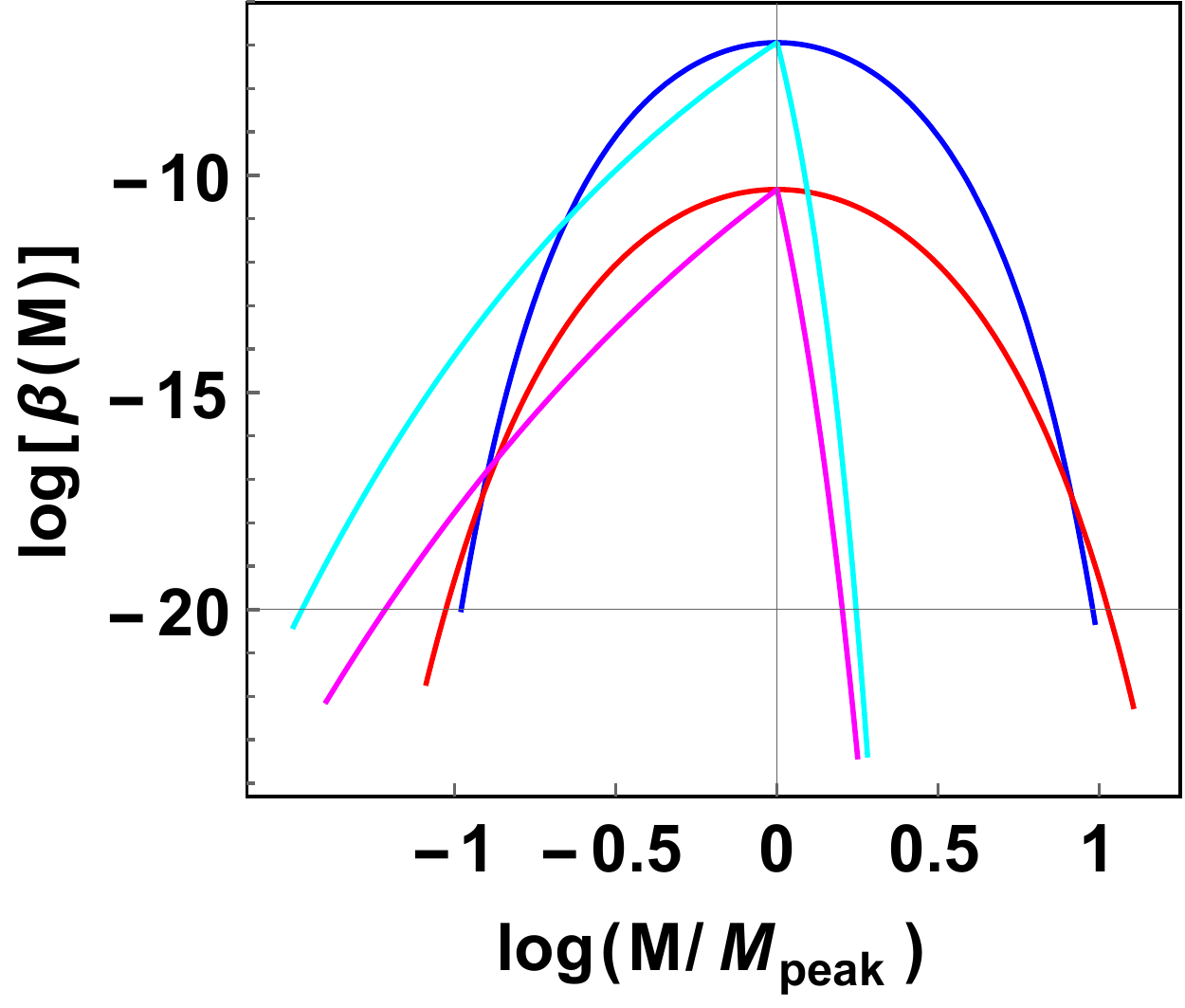}
\caption{$\beta (M)$ plotted for the Gaussian and Broken power law power spectra given in Equations~\eqref{powG} and~\eqref{powB}, respectively. Red and magenta curves are for $w=1/3$ for GPS and BPS, respectively. Blue and cyan curves are for $w=1$ for GPS and BPS, respectively. $P_0=0.02$ has been used for all the cases~presented.}
\label{betaKDM}
\end{center}
\end{figure}

\begin{center}
\begin{figure}[H]
%\begin{adjustwidth}{-\extralength}{0cm}
\centering %% If there is a figure in wide page, please release command \centering
\includegraphics[width=0.49\textwidth]{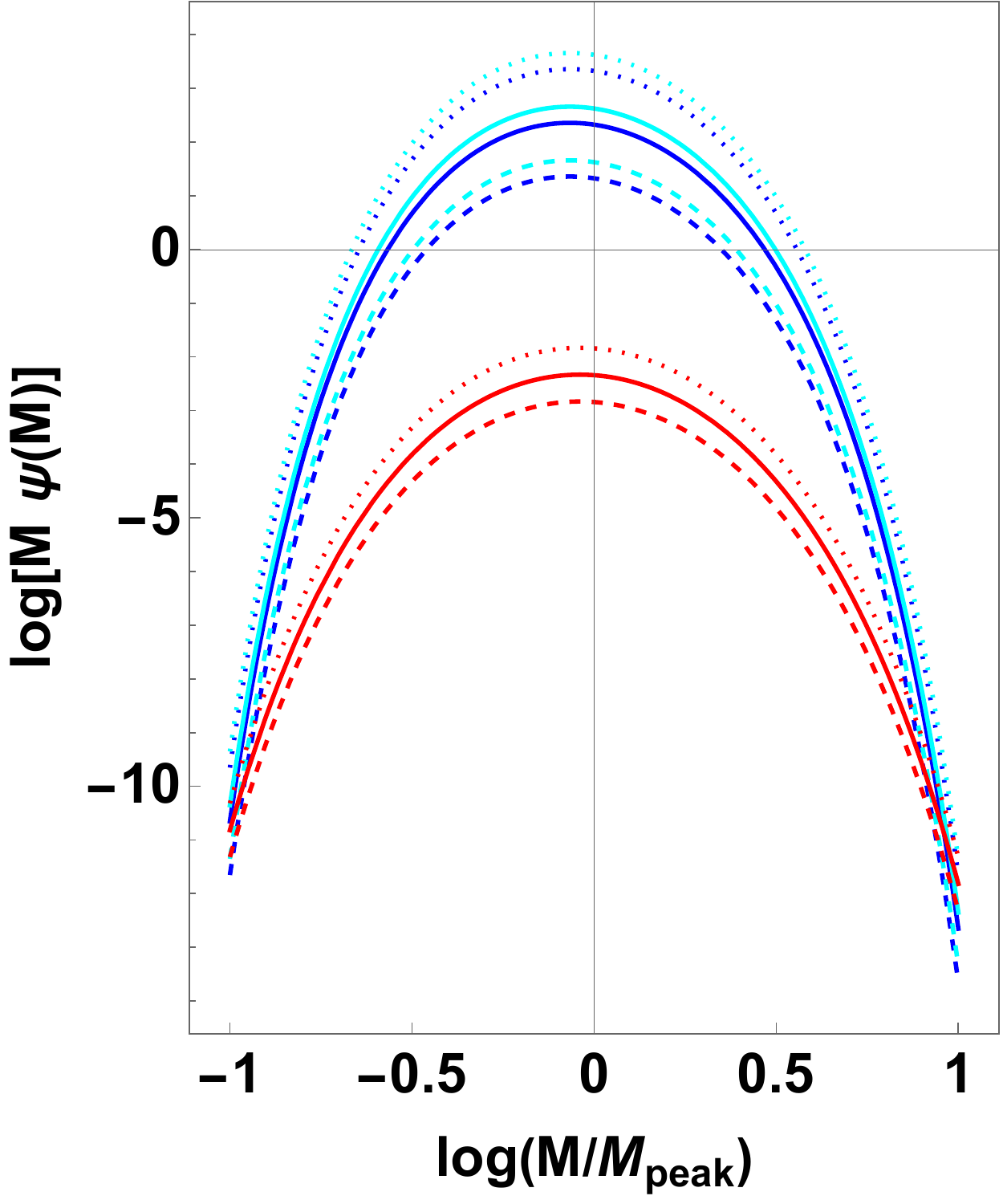}
\includegraphics[width=0.49\textwidth]{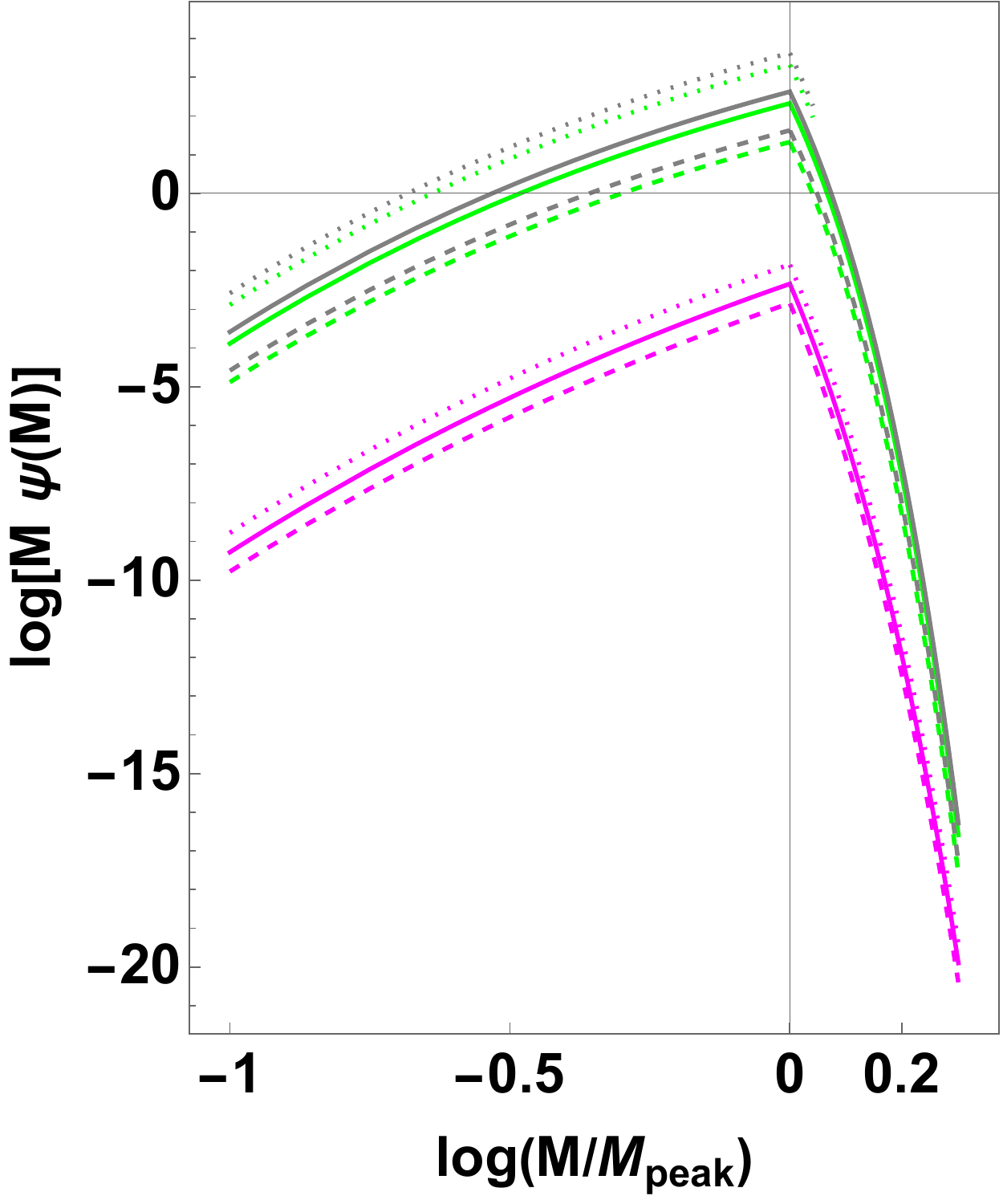}
%\end{adjustwidth}
\caption{The weighted mass function $M\psi (M)$ plotted for the Gaussian and Broken power law power spectra in the left and right panels, respectively. $P_0=0.02$ has been used for all the cases presented. In the left (right) panel, blue (green) curves are for $w=1$ with $T_{\rm RD}= 10$ MeV, cyan (grey) curves are for $w=1$ with $T_{\rm RD}= 5$ MeV and red (magenta) curves are for $w=1/3$. The solid, dashed and dotted lines signify the cases with $M_{\rm peak}=1$, $10$ and $0.1 M_{\odot}$, respectively.}
\label{psiKDM}
\end{figure}
\end{center}

In the following, it is shown that, for the same value of the peak amplitude of the primordial power spectra, $P_0=0.02$, PBH abundance in a KD epoch is more than that in a RD epoch. 
In Figure~\ref{betaKDM}, the mass fraction $\beta (M)$ is plotted for the KD and RD epochs for GPS and BPS as a function of the PBH mass normalized with $M_{\rm peak}=M(k_{\rm peak})$. From Equation~\eqref{Mkexactsol}, one can find that $M/M_{\rm peak}=(k/k_{\rm peak})^{-\frac{3(1+w)}{1+3w}}$; thus, the power spectra for GPS and BPS can be described only in terms of $\kappa =k/k_{\rm peak}$, without needing to specify $k_{\rm peak}$. Considerable improvement in the mass fraction for the KD case can be seen here. $\beta (M)$ is larger in the KD case than the RD case for a range of (0.1--10)$ M_{\rm peak}$ for the GPS, whereas for BPS, this range is (0.005--2)$ M_{\rm peak}$ (outside the range of the plot).
Figure~\ref{psiKDM} shows the improvement in the weighted mass function $M\psi (M)$ for the same scenarios, but with specific values of $M_{\rm peak}$, since $M\psi (M)\propto M^{\frac{-2w}{1+w}}$. The plots here are for $M_{\rm peak}=1$,~$10$ and $0.1 M_{\odot}$; however, similar improvements in $M\psi (M)$ can be seen for other peak masses as well. Here, the results are shown for two cases for the transition from the KD to RD epoch, $T_{\rm RD}=10$ MeV (blue and green curves) and $T_{\rm RD}=5$ MeV (cyan and grey curves). It can be seen that, when a PBH of a particular mass is produced, a lower value of $T_{\rm RD}$ leads to larger PBH abundance, which is expected from the dependence $\psi(M)\propto T_{\rm RD}^{-1}$ in Equation~\eqref{MpsiMw} for $w=1$.
%For both KD and RD, the total abundance $f_{\rm PBH}$ reached for each of the cases shown in Figure~\ref{psiKDM} is $\sim 10\%$.

%\unskip

In~\cite{Bhattacharya:2019bvk}, $M\psi (M)$ for the KD and RD epochs is compared for two different values of $k_{\rm peak}$ leading to abundant PBH around $M_{\odot}$ and $M\simeq 10^{18}$ gm. PBHs of mass around these two specific masses are of great interest since the former is of the order of black hole masses observed in binary mergers in LIGO/Virgo observations, whereas for the latter case, the possibility to attain $100\%$ of DM as PBHs is still not ruled out by observational bounds. Further results about PBH formation in a KD epoch can be found in~\cite{Bhattacharya:2019bvk} including the primordial amplitudes required for reaching $\sim$$10\%$ PBH abundance in DM, as well as the relevant modifications for IGW formed in a KD epoch.

\subsection{Early Matter Dominated Epoch}
\label{pbhsecmd}
An early epoch of matter domination can occur when a heavy field dominates the energy density for some time (see discussion in Section~\ref{genw1}). 
A well-studied example is moduli domination (mD) after inflation. Moduli is a scalar field $\Phi$ which at the end of inflation is frozen at its initial value $\Phi _0$. It starts moving in the potential once the Hubble parameter is such that $H\simeq m_{\Phi}$. Then, it keeps oscillating about the minimum of its potential, and the energy density carried by the field redshifts as matter ($a^{-3}$). This energy density dilutes slower than radiation and thus, at some time $T=T_*$, the energy density of the moduli starts to dominate the universe, marking the onset of mD. Finally, at $T=T_{\rm RD}$, the moduli decay (assuming instantaneous decay here) into visible and dark sector particles to produce a thermal bath of temperature that is suitable for BBN. Typically, the decay width $\Gamma _{\Phi}$ of a moduli field is given by 
\begin{equation}
\Gamma _{\Phi}=\frac{m_{\Phi}^3}{16\pi M_P^2}.
\end{equation}

During mD, $H_{\rm mD}=m_{\Phi}(\Phi _0/M_P)^4$, and the moduli field decays when $\Gamma _{\Phi}=H(T_{\rm RD})$, requiring $T_{\rm RD}> T_{\rm BBN}$. Thus, 
\begin{equation}\label{eq:trh1}
		T_{\rm RD}= \left(\dfrac{90}{\pi^2 g_*(T_{\rm RD})}\right)^{1/4} \sqrt{\Gamma _{\Phi}M_{P}} = 2.75{ ~MeV~}\bigg(\frac{10.66}{g_*(T_{\rm RD})}\bigg)^{1/4}\bigg(\frac{m_{\Phi}}{100 TeV}\bigg)^{3/2}\,
	\end{equation}
	
The bound from BBN temperature translates to a bound on the moduli mass $m_{\Phi}\gtrsim 135$ TeV. If $m_{\Phi}=500$ TeV, then the transition from mD to RD occurs (assumed to be instantaneous) at $T_{\rm RD}\simeq 30$ MeV. 

PBH formation in a mD epoch is explored in~\cite{Bhattacharya:2021wnk}, where it is shown that, even though PBHs of mass 0.1--10$ \ms$ can be produced in abundance in a mD epoch lasting up to $T_{\rm RD}=4.3$ MeV, they can explain only a few of the events in LIGO/Virgo observations, and can only contribute to $\sim$$4\%$ of total DM abundance. If PBHs are formed due to the overdensities entering in this $w=0$ mD epoch, then the resulting modification in $\beta (M)$ and $\psi (M)$ as compared to RD formation of PBH can be evaluated using Equations~\eqref{betaMGauss},~\eqref{betaMD} and~\eqref{psiM0}. Here, the results are also shown in terms of $M_{\rm peak}=M(k_{\rm peak})$.
For the mD epoch, $P_0=5\times 10^{-3}$ has been used to show the results, whereas, for RD, $P_0$ is the same as before. 
\begin{figure}[H]
\begin{center}
\includegraphics[width=0.5\textwidth]{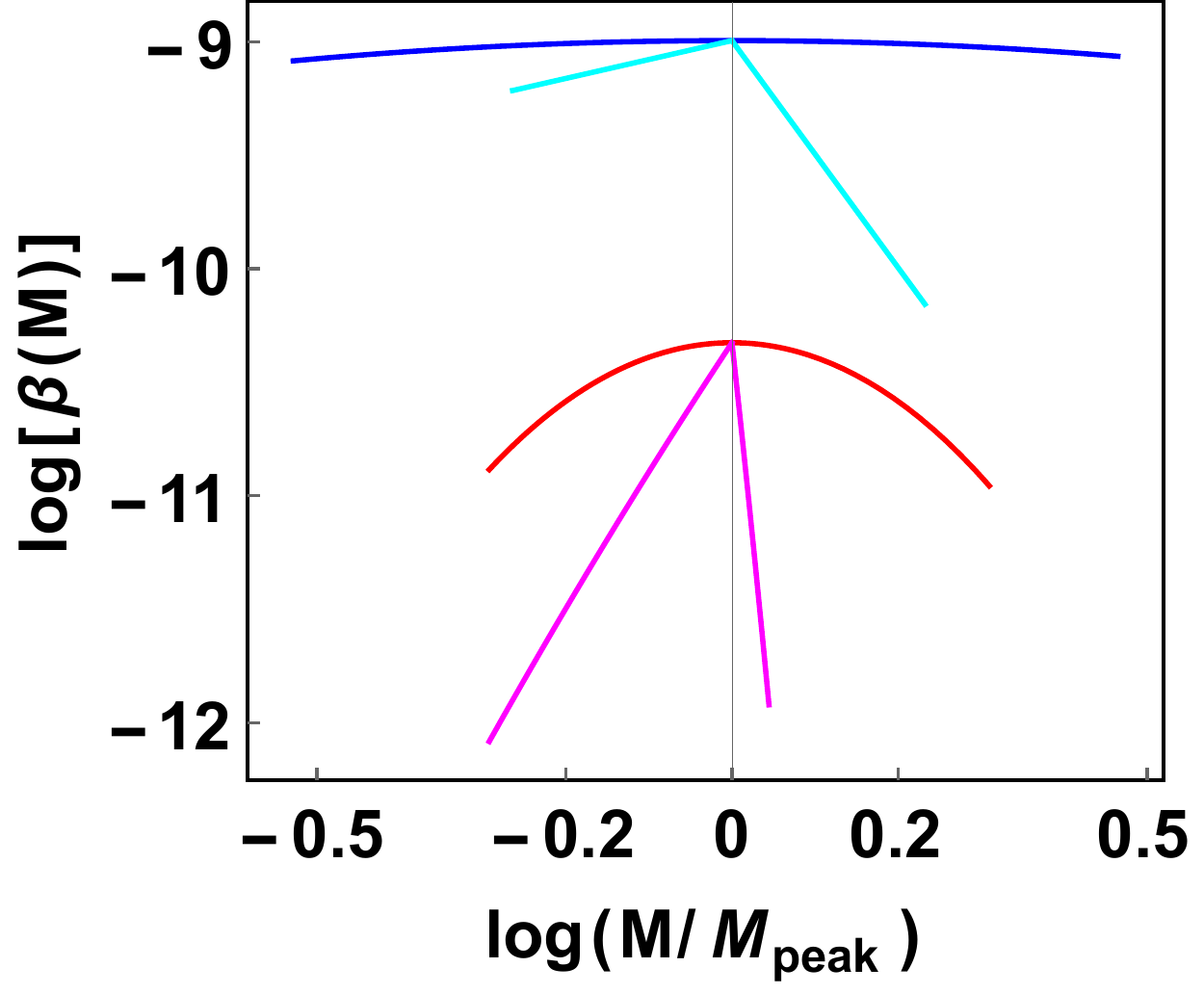}
\caption{$\beta (M)$ plotted for the Gaussian and Broken power law power spectra given in Equations~\eqref{powG} and~\eqref{powB} respectively. Red and magenta curves are for $w=1/3$ for GPS and BPS, respectively. Blue and cyan curves are for $w=0$ for GPS and BPS, respectively. $P_0=0.02$ for $w=1/3$ and $P_0=5\times 10^{-3}$ for $w=0$ have been used.}
\label{betaMDM}
\end{center}
\end{figure}
%\end{center}
%\begin{center}
\begin{figure}[H]
%\begin{adjustwidth}{-\extralength}{0cm}
\centering 
\includegraphics[width=0.49\textwidth]{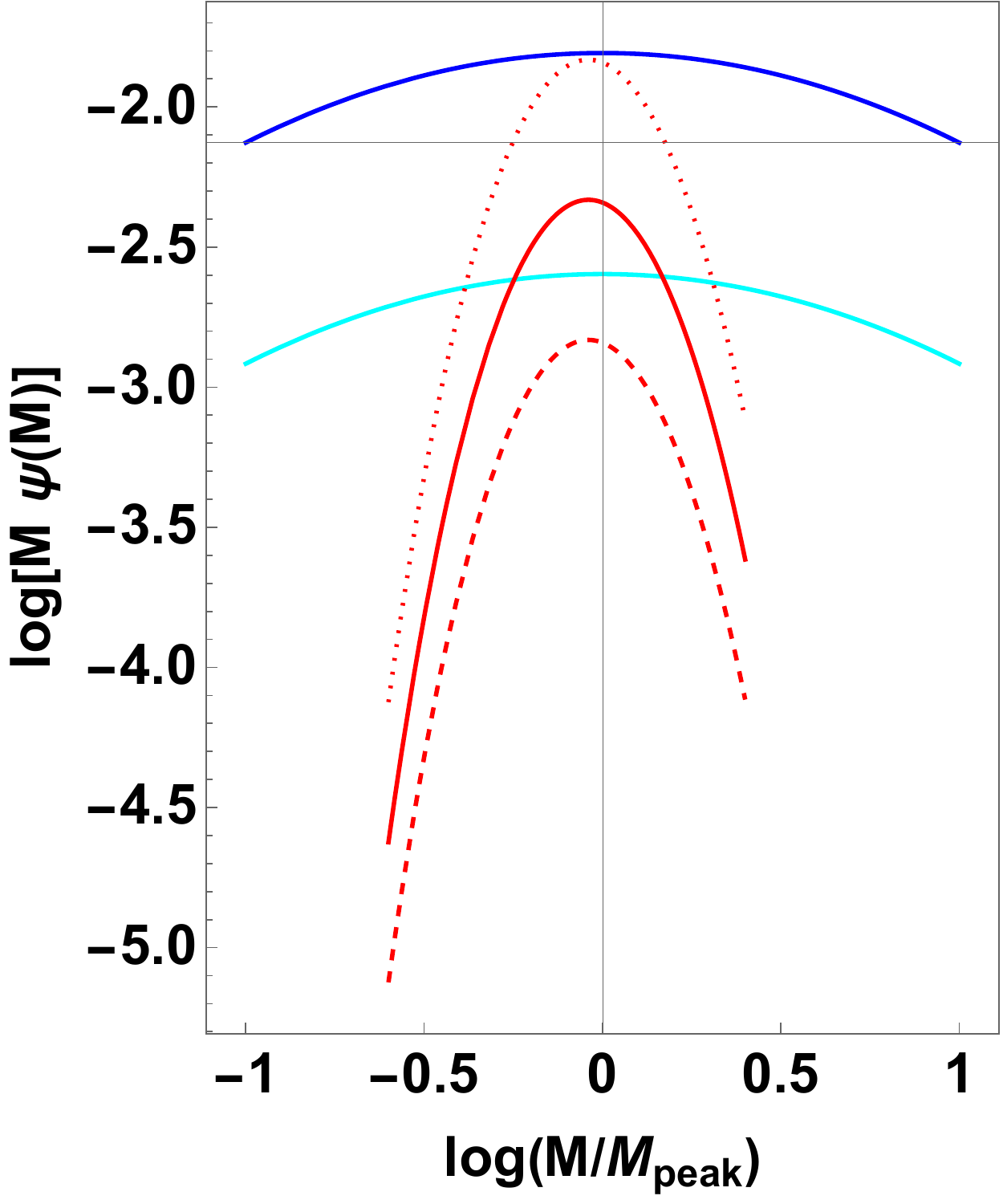}
\includegraphics[width=0.49\textwidth]{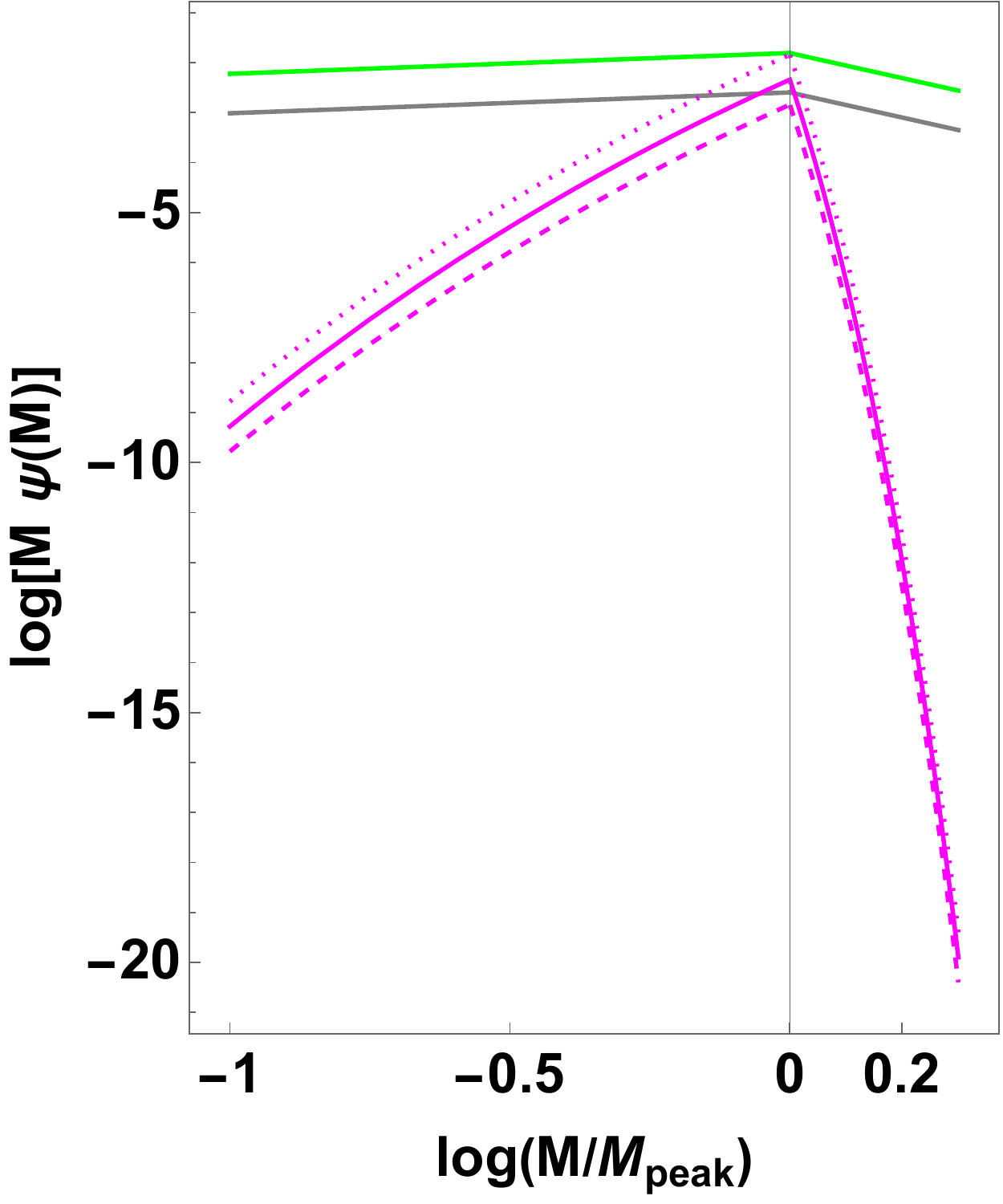}
%\end{adjustwidth}
\caption{The weighted mass function $M\psi (M)$ plotted for the Gaussian and Broken power law power spectra in the left and right panels, respectively. $P_0=0.02$ for $w=1/3$ and $P_0=5\times 10^{-3}$ for $w=0$ have been used. In the left (right) panel, the blue (green) curve is for $w=0$ with $T_{\rm RD}= 30$ MeV, the cyan (grey) curve is for $w=0$ with $T_{\rm RD}= 5$ MeV, and the red (magenta) curves are for $w=1/3$. The solid, dashed and dotted lines signify the cases with $M_{\rm peak}=1$, $10$ and $0.1 M_{\odot}$, respectively.}
\label{psiMDM}
\end{figure}
%\end{center}

In Figure~\ref{betaMDM}, the mass fraction $\beta (M)$ is plotted for the mD and RD epochs for GPS and BPS as a function of the PBH mass normalized with $M_{\rm peak}=M(k_{\rm peak})$. 
%, thus, the power spectra for GPS and BPS can be described only in terms of $\kappa =k/k_{\rm peak}$, without needing to specify $k_{\rm peak}$. 
Here, the improvement in the mass fraction for the mD case is evident; $\beta (M)$ is larger in the mD case than the RD case for a very large range due to the power law dependence of $\beta (M)$ on the primordial power spectrum (see Equation~\eqref{betaMDM}), even for a lower value of $P_0$ in the mD case.
%\begin{center}

Figure~\ref{psiMDM} shows the improvement in the weighted mass function $M\psi (M)$ for the same scenarios, but with specific values of $M_{\rm peak}$. However, for $w=0$, $M\psi (M)=M^0$; therefore, there is only one of each blue and cyan curve in this figure. These results are shown for two different transitions from the mD to RD epoch with $T_{\rm RD}= 30$ MeV (blue and green curves) and $T_{\rm RD}= 5$ MeV (cyan and grey curves)\footnote{$T_{\rm RD}= 5$ MeV corresponds to $m_{\Phi}=149$ TeV.}. Contrary to the KD case, here a lower value of $T_{\rm RD}$ leads to smaller PBH abundance, which can be understood from the dependence $\psi(M)\propto T_{\rm RD}$ in Equation~\eqref{MpsiMw} for $w=0$.
%For both mD and RD epochs, the total abundance $f_{\rm PBH}$ reached for each of the cases shown in Figure~\ref{psiMDM} is $\sim 10\%$.
%In reference~\cite{Bhattacharya:2021wnk} PBH formation in the epoch of moduli domination has been explored with the goal of finding a good abundance of near solar mass black holes, which can be (or already) observed in LIGO/Virgo surveys. It was found that very few of the 
%\begin{center}

It can also be seen explicitly that the peak amplitude of the primordial power spectrum in GPS required to produce a certain abundance of PBH is lower in both of the KD and mD cases as compared to the RD case. Here, Table~\ref{tablefpbh} shows the required values of $P_0$ to produce $10\%$ PBH abundances in these three epochs for two different values of $M_{\rm peak}$ with $T_{\rm RD}=5$ MeV for all the cases. As expected, the $P_0$ values required in KD and mD cases (last two columns) are lower than in the RD case (third column); however, for the case $M_{\rm peak}=10^{-12}M_{\odot}$, the $P_0$ required for mD is barely smaller than that required for the RD. This can be explained with the maximum mass that can be produced in a nonstandard epoch. For KD, the heaviest PBH of mass $M=1240M_{\odot}$ is produced when the perturbations enter the horizon at with $T_{\rm RD}=5$ MeV. However, for PBH formation in the mD epoch, 
as explained in the paragraphs before and after Equation~\eqref{MassHubEMD} in Section~\ref{formmd}, the heaviest mass produced for $T_{\rm RD}=5$ MeV is much smaller, $M\simeq 3M_{\odot}$. This sets upper limits for the integral in Equation~\eqref{ftot}, which affects the total~abundance. 

\begin{table}[H]\setlength{\tabcolsep}{6.8mm}
\caption{Necessary peak amplitude of GPS to reach $10\%$ PBH abundance for the specific non-standard post-inflationary scenarios discussed in this section.}
\label{tablefpbh}
\begin{tabular}{c cc c c}
\toprule
\boldmath{$M_{\rm peak}$}       & \boldmath{\textbf{$f_{\rm PBH}$} }& \boldmath{\textbf{$P_0$ for RD}} &  \boldmath{\textbf{$P_0$ for KD}} & \boldmath{\textbf{$P_0$ for mD}}                      \\ \midrule
%\hline
$M_{\odot}$ & $10\%$  & $0.0231$ & $0.0128$ & $0.0133$           \\ 
\midrule
%\hline
$10^{-12}M_{\odot}$ & $10\%$  & $0.0135$ & $0.0058$ & $0.0132$    \\\bottomrule
\end{tabular}
\end{table}
%\end{center}

\subsection{QCD Epoch}
During QCD phase transition around $T\simeq 200$ MeV, the strong interactions confine quarks into hadrons, while the effective number of relativistic dof changes rapidly. During this transition, thermodynamic quantities evolve smoothly, whereas the change in dof induces sudden dips in the EoS $w(T)$ and sound speed $c_s (T)$. Lattice QCD studies can deduce the evolution of $w(T)$ and $c_s^2 (T)$ during this transition. In~\cite{Byrnes:2018clq,Carr:2019kxo}, the change in the critical overdensity $\delta _c$ due to the change in $w(T)$ is derived. Dips in $w(T)$ and $c_s^2 (T)$ correspond to sudden transitions in  $\delta _c$ as much as from the usual RD value $0.453$ to a lower value $0.405$. Even this much change in $\delta _c$ can induce a large boost for PBH formation due to the exponential dependence of $\psi (M)$ on $\delta _c$. If a nearly scale-invariant density power spectrum enters the horizon during this time, then the PBH mass spectrum is boosted around the mass $M=\mathcal{O}(1)\ms$ (Ref.~\cite{Byrnes:2018clq} predicts the precise value of $M=0.7\ms$).

The idea of a softening of the EoS at particular energy scales has been extrapolated in~\cite{Carr:2019kxo} for the epochs when the pressure suddenly drops at $W^{\pm}/Z_0$ decoupling and during $e^+e^-$ annihilation, which resulted in boosting the PBH production for specific masses, which, interestingly, can explain some of the observed black holes in LIGO/Virgo surveys. Recently, Ref.~\cite{papanikolaou:2022cvo,Musco:2021sva} studied the $\delta _c$ determination for a dynamic $w$ and implemented this method to find a variation in $\delta _c(T)$, which is slightly different from previous studies with constant $w$~\cite{Escriva:2020tak}. 
%\subsection{Special Mentions}
%\section{Press--Schechter Method Vs Peak Theory}

\section{Discussions}
\label{conc}
The growing area of research on the topic of probing the early universe using PBHs is of utmost importance since it can shed light on both the small scales of inflation that are inaccessible to CMB surveys and the cosmological evolution before BBN. In this review, effects of possible non-standard epochs on PBH formation are discussed in detail with specific examples. The dependences of the main contributing quantities to the PBH abundance on the subtleties of model building and underlying assumptions have also been~emphasised.  

From an observational point of view, there are two main interesting aspects here: (i) surveys such as LIGO/Virgo may already have observed PBHs in the black hole merger events; (ii) some or all of the DM content in the universe can be explained with PBHs. For point (i), several propositions are made with particular inflation + post-inflation modelling to look for a good amount of PBH formation in the mass range consistent with the observed black holes in the LIGO/Virgo ``stellar graveyard''\footnote{See \url{https://www.ligo.caltech.edu/MIT/image/ligo20211107a} (accessed on 12/02/2023). }. For point (ii), various significant properties of PBH, such as lensing, Hawking radiation, etc., are used to provide upper bounds on the amount of PBHs of particular masses as DM. While developing a specific scenario of PBH formation, one checks the consistency of the predicted PBH abundance in DM with the observational bounds. This has been discussed more quantitatively, with specific examples of observational surveys in Section~\ref{importance}. In the same section, GWs induced by the large scalar fluctuations necessary for PBH formation are also discussed. Checking the consistency of predicted IGWs in different models with current and prospective GW surveys leads to interesting phenomenology since it can at least put upper bounds on the primordial power spectrum at relevant small scales. However, in the presence of a non-standard post-inflationary epoch, the amplitude and spectral shape of the IGW are also modified. The importance of combining PBH and IGW phenomenology, particularly in the presence of such non-standard cosmologies, is emphasised in this section. 

In Section~\ref{nonst}, general ideas about the origin of non-standard post-inflationary evolution have been put forward in the context of reheating epoch and additional epochs after instantaneous/slow reheating. In general, and as in this review, trivial assumptions are made, such as an instantaneous reheating epoch and an instantaneous transition from a non-standard $w$-dominated epoch to RD, but a realistic model of inflation is seldom that simple. Nevertheless, a lack of concrete understanding about inflationary reheating as well as a possible decay of additional dof after inflation (e.g., second reheating by a heavy field after its energy density dominated the universe with $w=0$) motivates one to present the general idea at first with such simple assumptions, and add complexities later on. However, works on PBH formation in slow reheating epochs and the exact evolution of $\delta _c$ during a dynamical EoS are interesting and very important, which reduce some of the uncertainties in specific cases. In this sense, Figure~\ref{evol} depicting the evolution of the scales of fluctuation and horizon will be modified for a realistic scenario with $\Delta N_{\rm rh}$ number of e-folds attributed to inflationary reheating and $(\Delta N_*, \Delta N_{\rm RD})$ e-folds attributed to the transition from RD to $w$-domination and back from $w$-domination to standard RD, respectively.
In Section~\ref{reason}, the cosmological evolution with the aforesaid assumptions has been discussed. Different examples of inflation models and the underlying mechanisms (e.g., inflection point) to result in growing $\mathcal{P}_{\zeta}(k)$ have also been referenced. 

In Section~\ref{analysis}, the main formalism for PBH formation in non-standard epochs has been developed. Firstly, the general $w$-dependent relation between PBH mass $M$ and cosmological scales has been derived. The appearance of the additional parameter $w$ here already hints at the modified relation between the inflationary sector ($\mathcal{P}_{\zeta}(k)$) and the PBH sector ($\psi (M)$). The PBH mass fraction $\beta (M)$ and mass function $\psi (M)$ have been developed for general $w$ as well as for the special case of $w=0$, i.e., an early matter dominated epoch. In a MD epoch, due to the complete absence of pressure, the process of  PBH formation is quite different and can incorporate interesting properties such as ellipticity, spin, etc. The basic quantities necessary to calculate the PBH abundance are the threshold of overdensity $\delta _c$, the distribution of overdensities $P(\delta )$, and initial mass fraction $\beta (M)$. Effects of nonlinearities, non-Gaussianities, and the shape of the fluctuation profile on these quantities have been discussed here, mentioning the simplified assumptions considered in this review.  

In Section~\ref{PBHspecific}, the formalism developed in Section~\ref{analysis} is applied for specific cases of non-standard evolution, namely kination $w=1$ and moduli domination $w=0$, and discussed for the softening of the EoS from the RD case during QCD transition. For all of these cases, PBH abundance is enhanced around the peak, which is shown for two different types of primordial power spectra (Equations~\eqref{powG} and~\eqref{powB}) for KD and mD cases. PBHs formed in the KD epoch can reach a higher abundance around the peak for the same order of peak amplitude as in RD, which is taken here to be $\mathcal{P}_0 = 0.02$. However, for mD, there is a gain in PBH abundance even for $\mathcal{P}_0=5\times 10^{-3}$ compared to $\mathcal{P}_0 = 0.02$ in RD. Mathematically, this improvement can be attributed to the power law relation between $\beta (M)$ and $\sigma (M)$ for mD as compared to the exponentially small dependence for $w>0$. 

The plots in this section are for specific values of $T_{\rm RD}=10$ MeV and $T_{\rm RD}=5$ MeV for KD and $T_{\rm RD}=30$ MeV and $T_{\rm RD}=5$ MeV for mD epochs, respectively. Naively, decreasing $T_{\rm RD}$ increases the PBH abundance further since the enhanced formation mechanism for $w\neq1/3$ sustains for a longer time. For $w>1/3$, this is evident from the dependence of $\psi (M)$ on $T_{\rm RD}$ in Equation~\eqref{psiM0} as $\psi (M)\sim T_{\rm RD}^{1-3w}$. For $w<1/3$, e.g., in MD, this dependence does not aid in enhancing abundance; however, the strong power law dependence between  $\beta (M)$ and $\sigma (M)$ again may come to the rescue to make the MD abundance of PBH more than the RD dominated one. However, as discussed before, for $w=0$, $\psi (M)\propto T_{\rm RD}$ means that ending an mD epoch later decreases the PBH abundance.  It should also be mentioned here that, in order to achieve $10\%$ abundance for both of the cases in Table~\ref{tablefpbh}, $P_0>0.01$, which does not strictly obey the condition for estimating the numerical results of~\cite{Harada:2016mhb} as the power law result for $\beta (M)$ in Equation \eqref{betaMD}. However, assuming that, by choosing a proper window function in Equation~\eqref{sigmaM1}, $\sigma <0.01$ can still be obtained with these values of $P_0$, the calculations are continued with the form in Equation~\eqref{betaMD}.

With the mechanism at hand, albeit with various simplified assumptions, it is high time to work with specific and concrete scenarios, leading to a combination of CMB consistent inflation models with growth in $\mathcal{P}_{\zeta}(k)$ at small scales and some duration of non-standard post-inflationary evolution. It is also of utmost importance to check the viability of the linear relation between the primordial fluctuations $\zeta$ and density perturbations $\delta $ for specific cases and incorporate primordial non-Gaussianities whenever necessary. If the uncertainties about (p)reheating and/or transition between $w$-dominated and RD epochs can be reduced for certain cases, then the predictions for PBH as well as IGW will be much more rigorous, which is hopeful for the verification of a particular scenario of the primordial universe with observations.

\vspace{6pt}

\acknowledgments{The research of S.B. is supported by the ``Progetto di Eccellenza'' of the Department of Physics and Astronomy of the University of Padua. S.B. also acknowledges the support from Istituto Nazionale di Fisica Nucleare (INFN) through the Theoretical Astroparticle Physics (TAsP) project. S.B. is also thankful to Suman Chatterjee, Anirban Das, Koushik Dutta, Subhendra Mohanty and Priyank Parashari for useful discussions. }

\newpage

%\reftitle{References}
%\begin{thebibliography}{999}
\bibliographystyle{utphys}
\bibliography{pbhmaster-2}

%\end{thebibliography}
%\PublishersNote{}
%\end{adjustwidth}
\end{document}